\documentclass[fleqn,usenatbib]{mnras}

\usepackage[T1]{fontenc}
\usepackage{ae,aecompl}
\usepackage{ulem}

\usepackage{graphicx}	
\usepackage{amsmath}	
\usepackage{amssymb}	

\newcommand{\app}[1]{Appendix~\ref{sec:#1}}
\newcommand{\eq}[1]{Equation~\ref{eq:#1}}
\newcommand{\fig}[1]{Figure~\ref{fig:#1}}
\renewcommand{\sec}[1]{Section~\ref{sec:#1}}
\newcommand{\tab}[1]{Table~\ref{tab:#1}}
\newcommand{\App}[1]{Appendix~\ref{sec:#1}}

\newcommand{\Fig}[1]{Figure~\ref{fig:#1}}
\newcommand{\Sec}[1]{Section~\ref{sec:#1}}
\newcommand{\Tab}[1]{Table~\ref{tab:#1}}

\newcommand{\bluetides}{\mbox{\sc{BlueTides}}}
\newcommand{\ceagle}{\mbox{\sc{C-Eagle}}}
\newcommand{\eagle}{\mbox{\sc{Eagle}}}
\newcommand{\lgals}{\mbox{\sc{L-Galaxies}}}
\newcommand{\euclid}{\mbox{\it Euclid}}
\newcommand{\flares}{\mbox{\sc Flares}}
\newcommand{\flare}{\mbox{\sc Flare}}

\newcommand{\romanst}{\mbox{\it Roman}}
\newcommand{\jwst}{\mbox{\it JWST}}
\newcommand{\SFR}{\psi}
\newcommand{\zoom}{\mbox{zoom}}

\newcommand{\Msun}{\mbox{M$_\odot$}}
\newcommand{\cMpch}{h$^{-1}$ cMpc}

\title[FLARES I]{First Light And Reionisation Epoch Simulations \\(FLARES) I: Environmental Dependence of High-Redshift Galaxy Evolution}

\author[C. C. Lovell et al.]{Christopher C. Lovell,$^{1,2}$\thanks{E-mail: c.lovell@herts.ac.uk (CCL)}
Aswin P. Vijayan,$^{2}$
Peter A. Thomas,$^{2}$
\newauthor
Stephen M. Wilkins,$^{2}$
David J. Barnes,$^{3}$
Dimitrios Irodotou,$^{2}$
Will Roper$^{2}$
\\
$^{1}$Centre for Astrophysics Research, School of Physics, Astronomy \& Mathematics, \\University of Hertfordshire, Hatfield AL10 9AB, UK\\
$^{2}$Astronomy Centre, University of Sussex, Falmer, Brighton BN1 9QH, UK\\
$^{3}$Department of Physics, Kavli Institute for Astrophysics and Space Research, Massachusetts Institute of Technology,\\Cambridge, MA 02139, USA
}

\date{Accepted XXX. Received YYY; in original form ZZZ}

\pubyear{2020}

\begin{document}
\label{firstpage}
\pagerange{\pageref{firstpage}--\pageref{lastpage}}
\maketitle

\begin{abstract}
We introduce the First Light And Reionisation Epoch Simulations (\flares), a suite of \zoom\ simulations using the \textsc{Eagle} model.
We resimulate a range of overdensities during the Epoch of Reionisation (EoR) in order to build composite distribution functions, as well as explore the environmental dependence of galaxy formation and evolution during this critical period of galaxy assembly.
The regions are selected from a large $(3.2 \;\mathrm{cGpc})^{3}$ parent volume, based on their overdensity within a sphere of radius $14$ \cMpch.
We then resimulate with full hydrodynamics, and employ a novel weighting scheme that allows the construction of composite distribution functions that are representative of the full parent volume.
This significantly extends the dynamic range compared to smaller volume periodic simulations.
We present an analysis of the galaxy stellar mass function (GSMF), the star formation rate distribution function (SFRF) and the star forming sequence (SFS) predicted by \flares, and compare to a number of observational and model constraints.
We also analyse the environmental dependence over an unprecedented range of overdensity.
Both the GSMF and the SFRF exhibit a clear double-Schechter form, up to the highest redshifts ($z = 10$).
We also find no environmental dependence of the SFS normalisation.
The increased dynamic range probed by \flares\ will allow us to make predictions for a number of large area surveys that will probe the EoR in coming years, carried out on new observatories such as \romanst\ and \euclid.
\end{abstract}




\section{Introduction}

A goal of numerical galaxy evolution studies is to model a representative population of galaxies, resolving all of the relevant physics at the required scales, in order to provide a test bed for the study and interpretation of observed galaxies \citep{benson_galaxy_2010}.
In order to achieve this it is necessary to simulate large volumes (in order to sample a representative volume of the Universe) at high resolution (\textit{e.g.} spatial, mass, time; in order to resolve the internal physical processes within individual galaxies) and with all of the key physics included (such as full hydrodynamics, magnetic fields, \textit{etc.}).
Unfortunately this is not computationally feasible; compromises must be made with volume, resolution or choice of physics, depending on the scientific questions posed \citep[for a review, see][]{somerville_physical_2015}.

The most common approach to obtain a representative population of galaxies is to simulate a large periodic cube, tens of Mpc across on a side.
This approach has been used in a number of leading projects to simulate large volumes down to redshift zero, producing thousands of galaxies across a wide range of stellar masses.
Projects such as \eagle\ \citep{schaye_eagle_2015,crain_eagle_2015}, \textsc{Simba} \citep{dave_simba:_2019}, Illustris \citep{vogelsberger_introducing_2014,genel_introducing_2014}, Illustris-TNG \citep{pillepich_first_2017,nelson_first_2017}, Romulus \citep{tremmel_romulus_2016} and Horizon-AGN \citep{dubois_dancing_2014} have mass resolutions of order $10^6 \mathrm{M_{\odot}}$, sufficiently high to resolve the internal structure of galaxies.
However, despite these large volumes, the rarest peaks of the overdensity distribution are still poorly sampled due to the lack of large scale modes in constrained periodic volumes.
Much larger volumes are required to sample the rare overdensities on large scales that are likely to evolve in to the most massive clusters by the present day.
For example, the \eagle\ simulation contains just 7, relatively low-mass clusters ($M_{200,c} > 10^{14} \, \mathrm{M_{\odot}}$) at $z=0$ within the fiducial 100 Mpc volume \citep{schaye_eagle_2015}.

One means of overcoming the limitations of relatively small periodic volumes is to use much larger, dark matter-only simulations, with box lengths of order Gpc, as sources for \zoom\ simulations.
These use regions selected from the dark matter only simulation as source initial conditions, and resimulate them at higher resolution with extra physics, such as full hydrodynamics \citep{katz_hierarchical_1993, tormen_structure_1997}.
This technique preserves the large scale power and tidal forces by simulating the dark matter at low resolution outside the high resolution region.
A recent example is the \ceagle\ simulations, high-resolution hydrodynamic simulations of 30 clusters with a range of descendant masses \citep{barnes_cluster-eagle_2017,bahe_hydrangea_2017}.
These were selected from a parent dark matter simulation with volume $(3.2 \;\mathrm{cGpc})^{3}$ \citep{barnes_redshift_2017}.
This enormous volume contains $185\,150$ clusters ($\mathrm{M_{200,c}} > 10^{14} \, \mathrm{M_{\odot}}$) and $1701$ high-mass clusters ($\mathrm{M_{200}} > 10^{15} \, \mathrm{M_{\odot}}$).
The \ceagle\ \zoom\ approach allowed the application of the \eagle\ model to cluster environments, without having to simulate a large periodic box.

The zoom technique can also be used to sample a range of overdensities, not just the peaks of the overdensity distribution.
The \textsc{GIMIC} simulations \citep{crain_galaxies-intergalactic_2009} are one example of this approach;
they picked 5 different regions of radius $20$ \cMpch\, at $z = 1.5$ from the Millennium simulation \citep{springel_simulations_2005}, with overdensities (-2,-1,0,1,2)$\sigma$ from the cosmic mean at z = 1.5.\footnote{where $\sigma$ is the \textit{rms} mass fluctuation on the resimulation scale}
These were then resimulated at high resolution with full hydrodynamics.
This not only allowed the investigation of the environmental effect of galaxy evolution, without having to simulate a whole periodic box, but also, by appropriately weighting each region according to its overdensity, the regions could be combined to produce composite distribution functions.
These composite functions have much larger dynamic range than those obtained from smaller periodic boxes, and at much lower computational expense than running a large periodic volume.

In this paper we use a similar approach to GIMIC to produce composite distribution functions of galaxy intrinsic properties, but focused on the Epoch of Reionisation (EoR).
The EoR approximately covers the first 1.2 billion years of the Universe's history ($4 \leqslant z \leqslant 15$), from the birth of the first Population III stars, to when the majority of the intergalactic medium is ionized \citep{bromm_first_2011,zaroubi_epoch_2013,stark_galaxies_2016,dayal_early_2018,cooray_cosmic_2019}.
A number of surveys over the past 15 years, with both space- and ground-based observatories, have discovered thousands of galaxies during this epoch \citep{beckwith_hubble_2006,warren_united_2007,wilkins_new_2011,koekemoer_candels_2011,grogin_candels:_2011,mccracken_ultravista:_2012,bouwens_uv_2015}.
Using intervening clusters as gravitational lenses has pushed the measurement of luminosity functions to even fainter magnitudes \citep{castellano_constraints_2016,livermore_directly_2017,atek_extreme_2018,ishigaki_full-data_2018}.
Spectral energy distribution fitting has been used to characterise the intrinsic properties of these galaxies, measuring for example their stellar masses \citep[\textit{e.g.}][]{gonzalez_evolution_2011,duncan_mass_2014,song_evolution_2016,stefanon_rest-frame_2017} and star formation rates \citep[\textit{e.g.}][]{smit_star_2012,katsianis_evolution_2017}.
However, we have yet to unambiguously detect Population III stars \citep{yoshida_formation_2019}, and the first stages of galaxy assembly are yet to be probed, particularly the seeding and early growth of super massive black holes \citep{smith_first_2017}.

However, this situation may soon change with the introduction of a number of new observatories, each with unique capabilities for exploring the EoR.
\jwst\ will provide unprecedented sensitive imaging with NIRCam, and follow up spectroscopy with NIRSpec and MIRI, to detect and characterise potentially the very first forming galaxies in the Universe \citep{gardner_james_2006}.
In tandem, \romanst\ and \euclid\ will produce wide-field surveys of the EoR \citep{spergel_wide-field_2015,laureijs_euclid_2011}.
These surveys will predominantly probe the bright end of the rest-frame UV Luminosity Function (UVLF), which is currently poorly constrained by periodic hydrodynamic simulations due to their small volume.
They will also discover some of the most extreme galaxies, in terms of luminosity and intrinsic mass, in the observable Universe at these redshifts \citep{behroozi_most_2018}.
These observations will be important to constrain models of galaxy formation and evolution, but it is also possible to predict observed populations in advance and test the recovery of intrinsic parameters \citep{pforr_recovering_2012,pforr_recovering_2013,smith_deriving_2015,lower_how_2020}.

Predictions for upcoming wide-field surveys have so far typically been made using phenomenological models.
One such class of methods are Semi-Analytic Models (SAMs), run on halo merger trees extracted from dark matter-only simulations \citep[for a review, see][]{baugh_primer_2006}.
Due to their efficiency they can be applied to large cosmological volumes, and used to probe distribution functions of intrinsic properties and observables over a large dynamic range.
A number of these models have been tested during the EoR \citep{henriques_galaxy_2015,clay_galaxy_2015,somerville_star_2015, poole_dark-ages_2016, rodrigues_constraints_2017, yung_semi-analytic_2019,lagos_far-ultraviolet_2019}.
Mock observables can also be produced and directly compared with observed luminosity functions \citep{lacey_unified_2016,yung_semi-analytic_2019-1,vijayan_detailed_2019}.
Such models can be run relatively quickly, allowing parameter estimation through Monte Carlo approaches \citep{henriques_galaxy_2015}, a powerful means of exploring large degenerate parameter spaces.
However, despite recent progress in resolving SAM galaxies in to multiple components \citep[\textit{e.g.}][]{henriques_l-galaxies_2020}, such models necessarily do not self-consistently model physical processes on small scales, relying on analytic recipes.

\begin{figure}
	\includegraphics[width=0.5\textwidth]{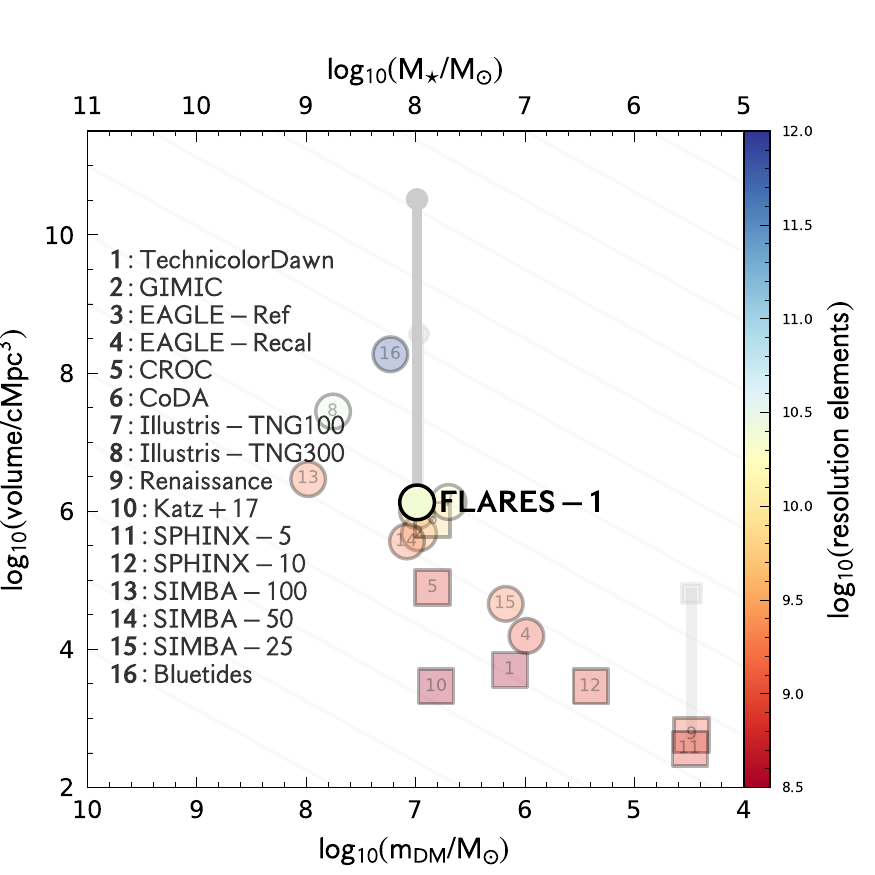}
    \caption{Dark matter element resolution against simulated volume.
		The colour of individual points describes the approximate number of resolution elements (dark matter + baryonic gas, excluding stars).
    We show the following simulation projects: Technicolor Dawn \citep{finlator_reionization_2018}, GIMIC \citep{crain_galaxies-intergalactic_2009}, EAGLE \protect\citep{schaye_eagle_2015,crain_eagle_2015}, CROC \protect\citep{gnedin_cosmic_2014}, CoDa \protect\citep{ocvirk_cosmic_2016}, Illustris \protect\citep{vogelsberger_introducing_2014}, Renaissance \protect\citep{barrow_first_2017}, the \protect\cite{katz_interpreting_2017} simulations, SPHINX \protect\citep{rosdahl_sphinx_2018}, and \bluetides\ \protect\citep{feng_bluetides_2016}.
    We also show \flares\ with the total resimulated high-resolution volume, as well as a vertical line showing the representative volume, given by that of the parent box.
    There is a strong negative correlation for periodic volumes between the volume that can be simulated and the resolution that can be achieved.
    The resimulation approach, with appropriate weighting, allows us to extend the volume axis significantly.
		}
    \label{fig:volume_resolution}
\end{figure}

Most existing periodic hydrodynamic simulations during the EoR are not able to achieve the large dynamic ranges accessible by SAMs.
This is illustrated in \fig{volume_resolution}, which shows where a number of existing simulations lie on a plane of simulated volume against hydrodynamic element mass.
There is a strong negative correlation, with some outliers.
The \bluetides\ simulation \citep{feng_bluetides_2016,feng_formation_2015}, based on the Massive Black suite of simulations \citep{matteo_cold_2012,khandai_massiveblack-ii_2015}, was performed within a (500 \,/\,h\,cMpc)$^3$ periodic box, $\sim$125 times as massive as the fiducial \eagle\ volume, whilst at a similar resolution.
They make predictions for a number of intrinsic and observational properties during the EoR \citep[\textit{e.g.}][]{waters_forecasts_2016, di_matteo_origin_2017, wilkins_photometric_2016, wilkins_lyman-continuum_2016, wilkins_properties_2017, wilkins_dust-obscured_2018, wilkins_nebular_2020}
Unfortunately, due to the increased computational cost it has only been run down to $z = 7$, and the model cannot therefore be tested against low redshift observables.
Other simulations have taken a different approach, instead simulating smaller volumes at much higher resolution, allowing them to investigate the effect of a number of physical processes in greater detail \protect\citep{oshea_probing_2015,jaacks_legacy_2019}.
However, these must similarly be stopped at intermediate redshifts due to the higher computational expense.

In this paper we introduce \flares, zoom resimulations during the EoR using the \eagle\ model.\footnote{project website available at \url{https://flaresimulations.github.io/flares/}}
The \eagle\ project \citep{schaye_eagle_2015,crain_galaxies-intergalactic_2009} is a suite of Smoothed Particle Hydrodynamics (SPH) simulations, calibrated to reproduce the stellar mass function and sizes of galaxies in the local Universe.
\eagle\ has been shown to be in good agreement with a large number of observables not used in the calibration \citep[e.g.][]{lagos_molecular_2015,bahe_distribution_2016,furlong_size_2017,trayford_colours_2015,trayford_optical_2017,crain_eagle_2017}.
This includes predictions at high-redshift: \cite{furlong_evolution_2015} found reasonably good agreement with observationally inferred distribution functions of stellar mass and star formation rate out to $z = 7$.
Unfortunately, there are very few galaxies in the fiducial \eagle\ volume during the EoR.
This is particularly the case for the most massive objects, which predominantly reside in protocluster environments, the progenitors of today's collapsed clusters \citep{chiang_galaxy_2017,lovell_characterising_2018}.
\flares\ allows us to significantly increase the number of galaxies simulated during the EoR with \eagle.
It also allows us to test the already incredibly successful \eagle\ model in a new regime of extreme, high-$z$ environments, whilst still resolving hydrodynamic processes at $10^{6} \, M_{\odot}$ resolution, and provide predictions for a number of key upcoming observatories.

In this, the first \flares\ paper, we introduce the resimulation method, our suite of zoom simulations, and present our first predictions for the distribution of galaxies by stellar mass and star formation rate using the composite approach.
This is the first in a series of papers studying the galaxy properties in the \flares\ sample; in Paper II we forward-model the full spectro-photometric properties, and predict the UV luminosity function and its high redshift evolution (Vijayan et al., \textit{in prep.}).
We assume a Planck year 1 cosmology \citep[$\Omega_{0} = 0.307$, $\Omega_{\Lambda} = 0.693$, $h = 0.6777$, ][]{planck_collaboration_planck_2014} and a Chabrier stellar initial mass function (IMF) throughout \citep{chabrier_galactic_2003}, and have corrected observational results accordingly.

\section{The \flare\ Simulations}
\label{sec:method}

We will now detail our simulations, including the \eagle\ model, selection of the regions, the zoom resimulation technique, and our method for constructing composite distribution functions.

\begin{figure}
	\includegraphics[width=0.98\columnwidth]{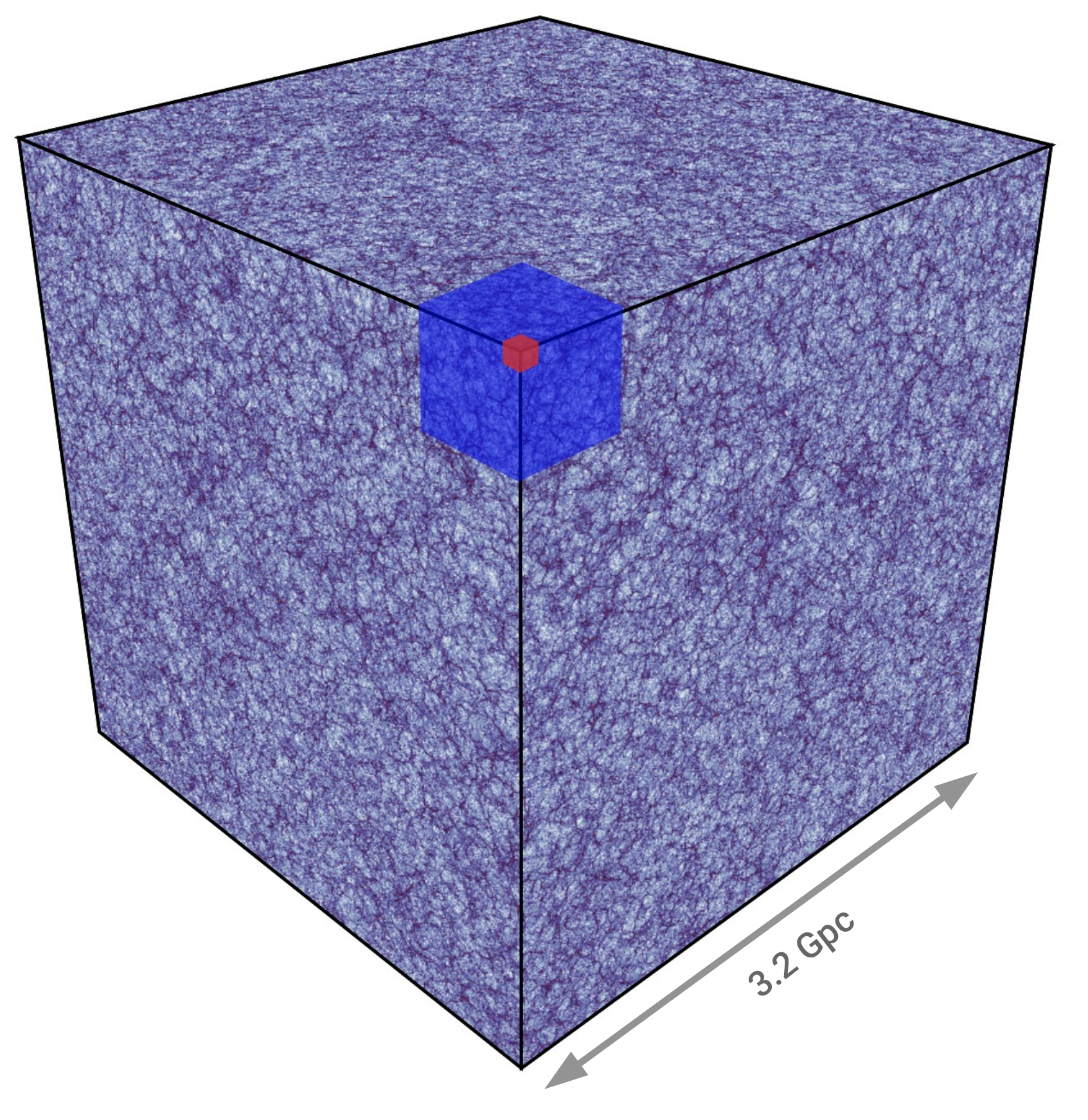}
    \caption{Diagram of the 3.2\,cGpc box from which we select our regions \citep{barnes_redshift_2017}.
		To demonstrate the increase in volume, we show the Bluetides simulation \protect\citep[$L = 570$\,cMpc;][]{feng_bluetides_2016} inset in blue, and the fiducial EAGLE simulation \protect\citep[$L = 100$\,cMpc;][]{schaye_eagle_2015} inset in red.}
    \label{fig:L3200}
\end{figure}

\begin{figure*}
  \includegraphics[width=\textwidth]{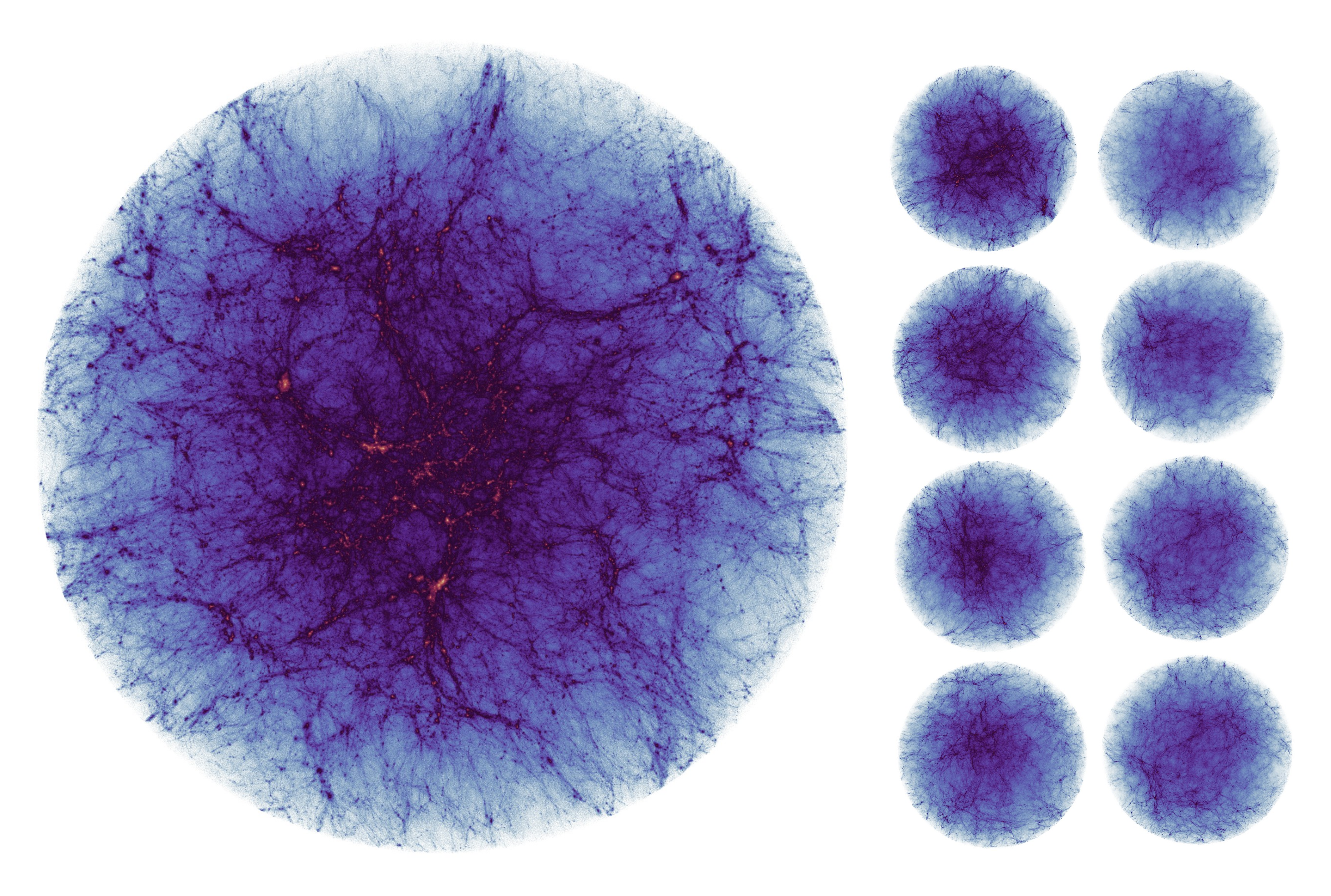}
  \caption{Visualisation of the dark matter integrated density in a number of resimulation regions of differing overdensity ($\delta$), made with \textsf{Py-SPHViewer} \citep{alejandro_benitez_llambay_2015_21703}.
	The region on the left shows the most overdense region ($00$, $\delta  = 0.970$).
	The regions to the right are (anticlockwise from top left) $17$, $20$, $22$, $24$, $26$, $28$, $30$, $38$, with overdensities $\delta = [0.616,0.266,0.121,-0.007,-0.121,-0.222,-0.311,-0.479]$, respectively.
	}
  \label{fig:sm_vis}
\end{figure*}

\subsection{The \eagle\ Model}

The \eagle\ physics model is based on that developed for the \textsc{OWLS} project \citep{schaye_physics_2010}, which is a heavily modified version of \textsc{P-Gadget-3} \citep{springel_simulations_2005}, an $N$-body tree-PM SPH code.
The hydrodynamics suite is collectively known as `\textsc{Anarchy}', (described in Appendix A of \cite{schaye_eagle_2015}, and \cite{schaller_eagle_2015}).
In short, it consists of the \cite{hopkins_general_2013} pressure-entropy SPH formalism, an artifical viscosity switch \citep{cullen_inviscid_2010}, an artificial conductivity switch \citep[e.g.][]{price_modelling_2008}, the \cite{wendland_piecewise_1995} $C^{2}$ smoothing kernel with 58 neighbours, and the \cite{durier_implementation_2012} time-step limiter.

Radiative cooling, formation of stars, black hole seeding, and feedback from stars and black holes are all handled by subgrid models.
Full details are provided in \cite{schaye_eagle_2015,crain_eagle_2015}.
We use the AGNdT9 parameter configuration, which produces similar mass functions to the reference model but better reproduces the hot gas properties in groups and clusters \citep{barnes_cluster-eagle_2017}.
This is identical to that used in the \ceagle\ simulations, but differs from the fiducial Reference simulation (see \tab{example_table}).
It uses a higher value for $C_{\mathrm{visc}}$, which controls the sensitivity of the BH accretion rate to the angular momentum of the gas, and a higher gas temperature increase from AGN feedback, $\Delta T$.
A larger $\Delta T$ leads to fewer, more energetic feedback events, whereas a lower $\Delta T$ leads to more continual heating.
These parameter changes impact the central black hole accretion, which has been shown to be efficient only at halo masses $> \, 10^{12} \, M_{\odot}$ \citep{bower_dark_2017}.
At $z = 10$ no \flares\ galaxies reside in such halos, however at $z = 5$ a minority do ($< 0.2 \%$), which may affect the early star formation histories of cluster galaxies \citep{bahe_hydrangea_2017}.
We use an identical resolution to the fiducial \eagle\ simulation, with gas particle mass $m_{g} = 1.8 \times 10^6 \; \mathrm{M_{\odot}}$, and a softening length of $2.66 \; \mathrm{ckpc}$.

\begin{table}
	\centering
	\caption{Variation of subgrid parameters between models. }
	\label{tab:example_table}
	\begin{tabular}{lcccc}
		\hline
		\textbf{Simulation Prefix} & $C_{\mathrm{visc}}$ & $\Delta T_{\mathrm{AGN}}$\\
		\hline
      & & $\mathrm{[K]}$ \\
		\hline
		Ref & $2 \pi$ & $10^{8.5}$ \\
		AGNdT9 & $2 \pi \times 10^{2}$ & $10^{9}$ \\
		\hline
	\end{tabular}
\end{table}

\subsection{Region Selection}
\label{sec:method:region}

We use the same parent simulation as that used in the \ceagle\ simulations \citep{barnes_redshift_2017}: a (3.2\,cGpc)$^3$ dark matter-only simulation with a particle mass of $8.01 \,\times\, 10^{10}$\,\Msun, using a \cite{planck_collaboration_planck_2014} cosmology.
\fig{L3200} shows a diagram of the box compared to the fiducial \eagle\ reference volume, as well as the \bluetides\ simulation \citep{feng_bluetides_2016}.
The highest redshift snapshot available for this simulation is at $z = 4.67$, which we use for our selection.
Within this snapshot, we select spherical volumes that sample a range of overdensities.
By taking a sufficiently large radius we can ensure that the density fluctuations averaged on that scale are linear, such that the distortion in the shape of the Lagrangian volume during the simulation will not be too extreme and that the ordering of the density fluctuations is preserved.
The regions, and their overdensities, are given in \tab{regions}.

\begin{figure}
  \includegraphics[width=\columnwidth]{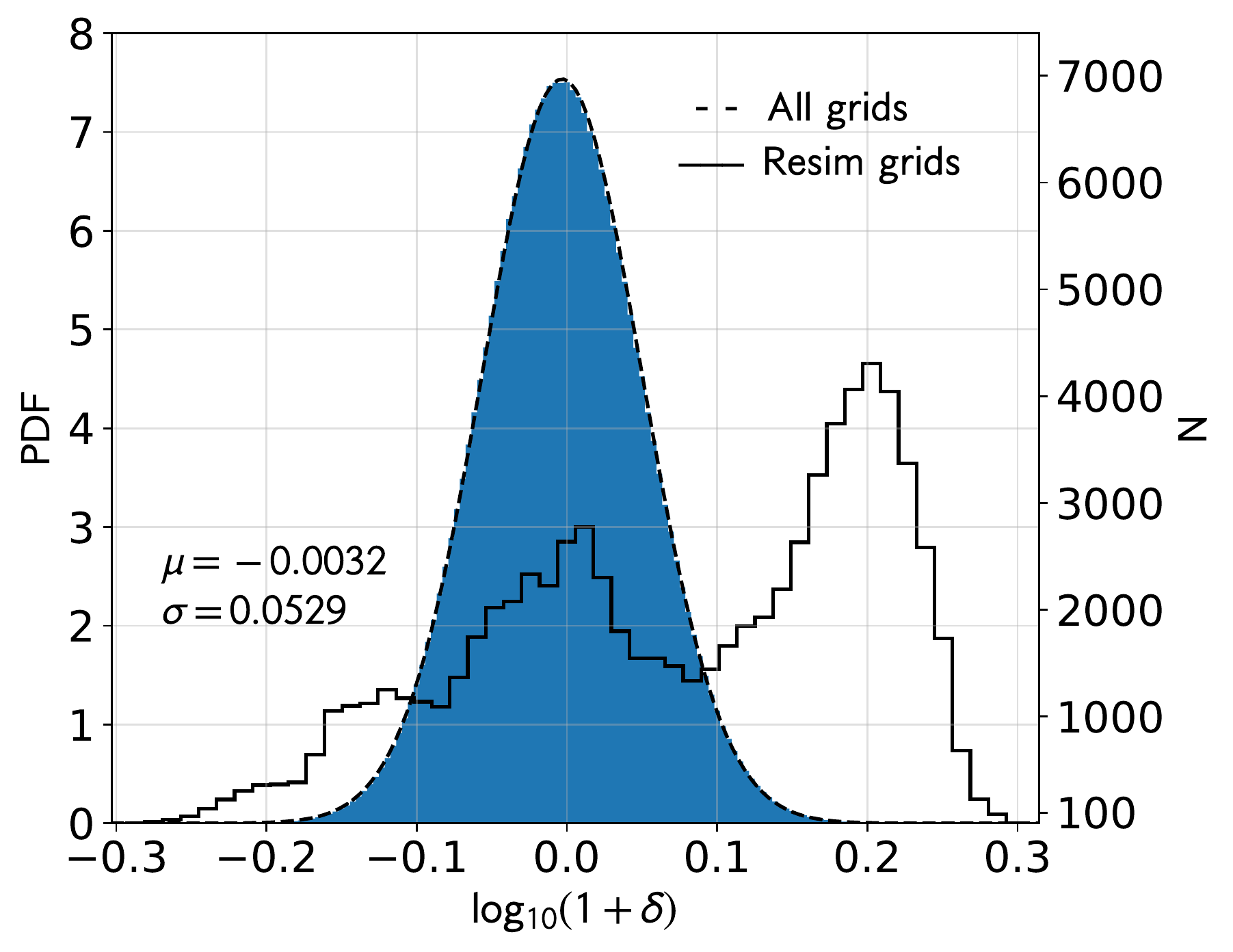}
  \hspace*{0.15cm}\includegraphics[width=0.85\columnwidth]{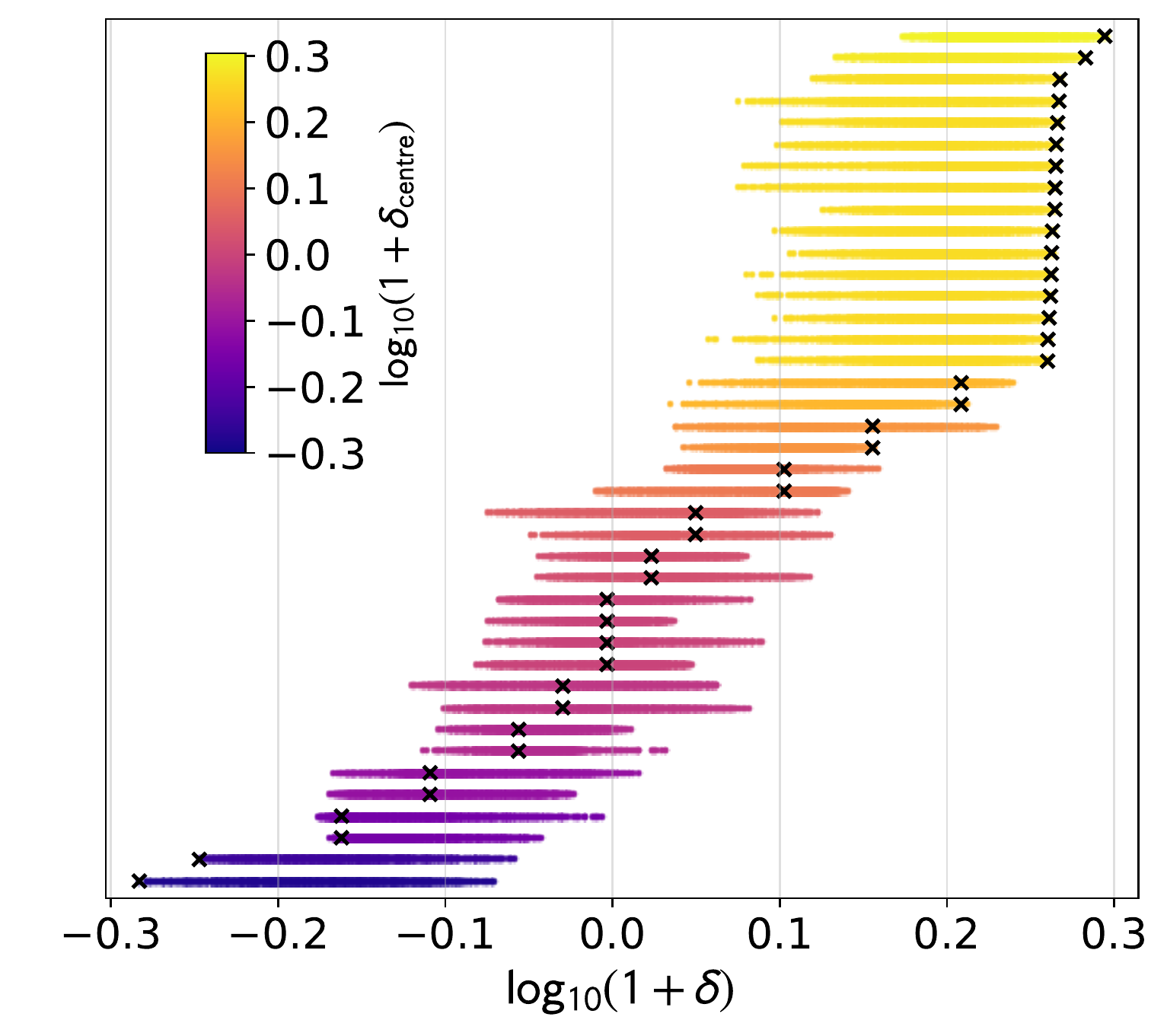}
  \caption{\textit{Upper panel}: the probability distribution function of sampled overdensities.
	The dashed black line shows a lognormal fit with the given parameters.
	The solid blue histogram shows the grid locations that lie within one of our resimulation volumes.
	The solid black histogram shows the distribution of our selected regions in overdensity, binned into 50 equal width bins, with the right y-axis showing only their number counts.
	\textit{Lower panel}: the distribution of overdensities within each simulation volume. The vertical displacement is arbitrary.  The cross shows the overdensity measured at the centre of the resimulated volume and the spread of values shows the overdensities within each volume evaluated at each point on the 2.67\,cMpc grid.}
  \label{fig:log_fit}
\end{figure}

To determine the density, we first distribute the mass onto a high resolution, $3.2 \; \mathrm{cGpc}\,/\,1200 \sim 2.67\; \mathrm{cMpc}$ cubic grid using a nearest grid point assignment scheme.
We then find the density on larger scales by convolving the grid with a spherical top-hat filter of radius $14$ \cMpch.\footnote{Code provided at \\\href{https://github.com/christopherlovell/DensityGridder}{https://github.com/christopherlovell/DensityGridder}}
We find, in test volumes, that this gives densities very close to those calculated from the raw particle data.
The overdensity is then defined as
\begin{equation}\label{eq:1}
	\delta(\textbf{x}) = \frac{\rho(\textbf{x})}{\bar\rho} - 1,
\end{equation}
where $\rho$ is the density at grid coordinates $\textbf{x}$, and $\bar\rho$ is the mean density in the box.
The upper panel of \fig{log_fit} shows the distribution of overdensity in log-space, alongside a fitted log-normal distribution.

We select regions for resimulation with two different goals: firstly, we select a number of regions of high overdensity in order to obtain a large sample of the first massive galaxies to form in the Universe; and secondly we select regions with a range of overdensities in order to explore the environmental impact (bias) on galaxy formation.
In order to achieve the first goal we select the 16 most overdense regions in the volume, which have $\delta \geqslant 0.8$.
For the second goal, we select two regions at each overdensity based on their \textit{rms} overdensity $\sigma$, in the range $\sigma \in [4,3,2,1,0.5,0,-0.5,-1,-2,-3]$.
We choose two regions of each overdensity in order to minimise the effect of cosmic variance at fixed overdensity; we also select an additional two mean density regions, to increase the sampled volume of these common regions.
Finally, we also select the two most underdense regions ($\delta \sim -0.45$) in order to cover the whole dynamic range.
This gives a total of 40 regions.
Figures \ref{fig:gsmf_overdensity} and \ref{fig:sfrf_overdensity} show the overdensity dependence of the GSMF and SFRF, respectively; at fixed overdensity the poisson noise is low, which suggests the effect of cosmic variance is low, and that the number of regions chosen was sufficient to demonstrate the trends presented in this article.
However, we plan to run a greater number of simulations to further reduce the noise above the knee of the stellar mass function; an advantage of the resimulation approach is that this can simply be achieved by running more simulations to increase the total simulated volume.

The selected regions are listed in \app{regions} and the range of overdensities that each covers (evaluated at each point on the 2.67\,cMpc grid enclosed by that volume) is shown in the lower panel of \fig{log_fit}.
We discuss how to combine the resimulations so as to obtain a representative sample of the whole Universe in \Sec{method:weighting}.

\subsection{The Resimulation Method}
\label{sec:method:resim}

Galaxies on the edge of the high resolution region will not be modelled correctly due to the presence of a pressureless boundary.
To avoid this we resimulate a region 15\,\cMpch\, in radius, and ignore all galaxies within 1\,\cMpch\, of the edge of the sphere in post-processing.\footnote{We have tested and found that our results are insensitive to changes ($\pm 0.5 \,\mathrm{cMpc}$) in the size of this boundary region.}
At higher redshift the Lagrangian high resolution region can deform, but we found that it is close to spherical out to the highest redshifts considered in this work ($z = 10$).
\fig{sm_vis} shows the dark matter distribution within the cutout radius for a range of resimulations of differing overdensity, at $z = 4.7$.
We also show the fiducial periodic \eagle\ volume to provide a visual comparison of the differing environments probed.

As in the standard \eagle\ analysis, structures are first found using a Friends-Of-Friends \citep[FOF,][]{davis_evolution_1985} finder, then split into bound substructures using the \textsc{Subfind} algorithm \citep{springel_populating_2001}.
\footnote{A number of galaxies identified by subfind are, on close inspection, `spurious' structures, which manifest as an unrealistic ratio between the stellar, gas or dark matter components \citep[see][for a discussion]{mcalpine_eagle_2016}.
These galaxies make up less than 0.1\% of all galaxies $> \, 10^{8} \, M_{\odot}$ at $z = 5$, and are typically low mass.
We use the following conditions to flag spurious galaxies: any subhalo with zero mass in the stellar, gas or dark matter components.
Once these galaxies have been identified, we remove them from the subfind catalogues, and add their particle properties to the parent `central' subhalo.}
Their properties are then defined using those stellar particles within 30\,pkpc of the location of the most tightly-bound stellar particle.
We limit our analysis to galaxies sampled by at least 50 star particles, which corresponds to a mass limit of approximately $\mathrm{log_{10}}(M_\star \,/\,\mathrm{M_{\odot}}) \geqslant 7.95$.

\subsection{Distribution Function Weighting}
\label{sec:method:weighting}

In this section we describe how we combine our resimulations to obtain a statistically-correct representation of the universal cosmological distribution of galaxies.
As we show below in \Sec{results:gsmf}, distribution functions, such as the galaxy stellar mass function, vary with the overdensity of the resimulated volume.
Therefore, it is necessary to \textit{weight} each resimulation to reproduce the correct distribution of those overdensities averaged over the whole Universe, i.e. the cosmic mean.

As mentioned in \Sec{method:region}, the overdensity within spherical top-hat regions of radius 14\,\cMpch\, is sampled on a 2.67\,cMpc grid; we label this sample $\delta_g$.
Since the grid sampling is finer than the size of the resimulation volume, each resimulation volume is associated with just under 2000 different values of $\delta_g$.
We show the distribution of those $\delta_g$ within each resimulation volume in the lower panel of \Fig{log_fit}.
The most overdense regions, whilst containing a single highly overdense point, in fact contain points covering a range of overdensities.
It is, therefore, important to account for this spread in sampled overdensity, rather than just using the central overdensity when determining the contribution from any particular resimulation volume.

The top panel of \Fig{log_fit} contrasts the PDFs of $\delta_g$ for the whole box and for our resimulated sample.
To generate the correct mean distibutions, we divide into bins of overdensity as shown by the histogram in \fig{log_fit} (black solid line), then \textit{weight} the resimulations appropriately to reproduce the cosmic distribution.
Specifically, we do the following:
\begin{itemize}
\item The overdensity domain is split up into 50 bins of equal width in $\log_{10}(1+\delta)$, $i=1 \ldots N_\delta$.\footnote{We tested using a greater number of bins and found that the quantitative weights did not change significantly.}
For each of these, it is possible to assign a weight, $w_{\mathrm{true},i}$, in proportion to the fraction of $\delta_g$ that lie in that bin, such that $\Sigma_iw_{\mathrm{true},i}=1$.
\item Each resimulation, $j$, is similarly distributed over these overdensity bins with weights, $w_{ij}$, in proportion to the enclosed values of $\delta_g$.  Thus $\Sigma_iw_{ij}=1$.
\item The sample weight associated with each bin is $w_{\mathrm{sample},i}=\Sigma_jw_{ij}$.
\item To obtain the correct universal average, we therefore have to weight each density bin by the ratio $r_i=w_{\mathrm{true},i}/w_{\mathrm{sample},i}$.
\end{itemize}
Ideally, we would associate each galaxy with the local value of $\delta_g$.
However, for the purposes of simplicity in this paper, we give all galaxies within a particular resimulation equal weight -- this will give some dispersion over the more correct method, which we will implement in a future paper.
\begin{itemize}
\item Hence we adjust the contribution of each resimulation by a factor $f_{j}=\Sigma_ir_iw_{ij}$.
\end{itemize}
We note that
\begin{align}
  \Sigma_j f_j &= \Sigma_j \Sigma_i r_i w_{ij} = \Sigma_i r_i \Sigma_j w_{ij} \nonumber \\
  &= \Sigma_i r_i w_{\mathrm{sample},i} = \Sigma_i w_{\mathrm{true},i} =1.
\end{align}
These simulation weighting factors are listed in \Tab{regions}.

We further note that, at higher redshifts, the overdensities will evolve.
Nevertheless, because even the most extreme perturbations are only mildly non-linear, we would expect that the ordering of the overdensities would largely be preserved.
Hence, we use the same sampling at all redshifts.
That also allows for a much more direct comparison of the evolution within each overdensity sample.

\section{Results}

\subsection{Galaxy Number Counts}

\begin{figure}
    \centering
    \includegraphics[width=\columnwidth]{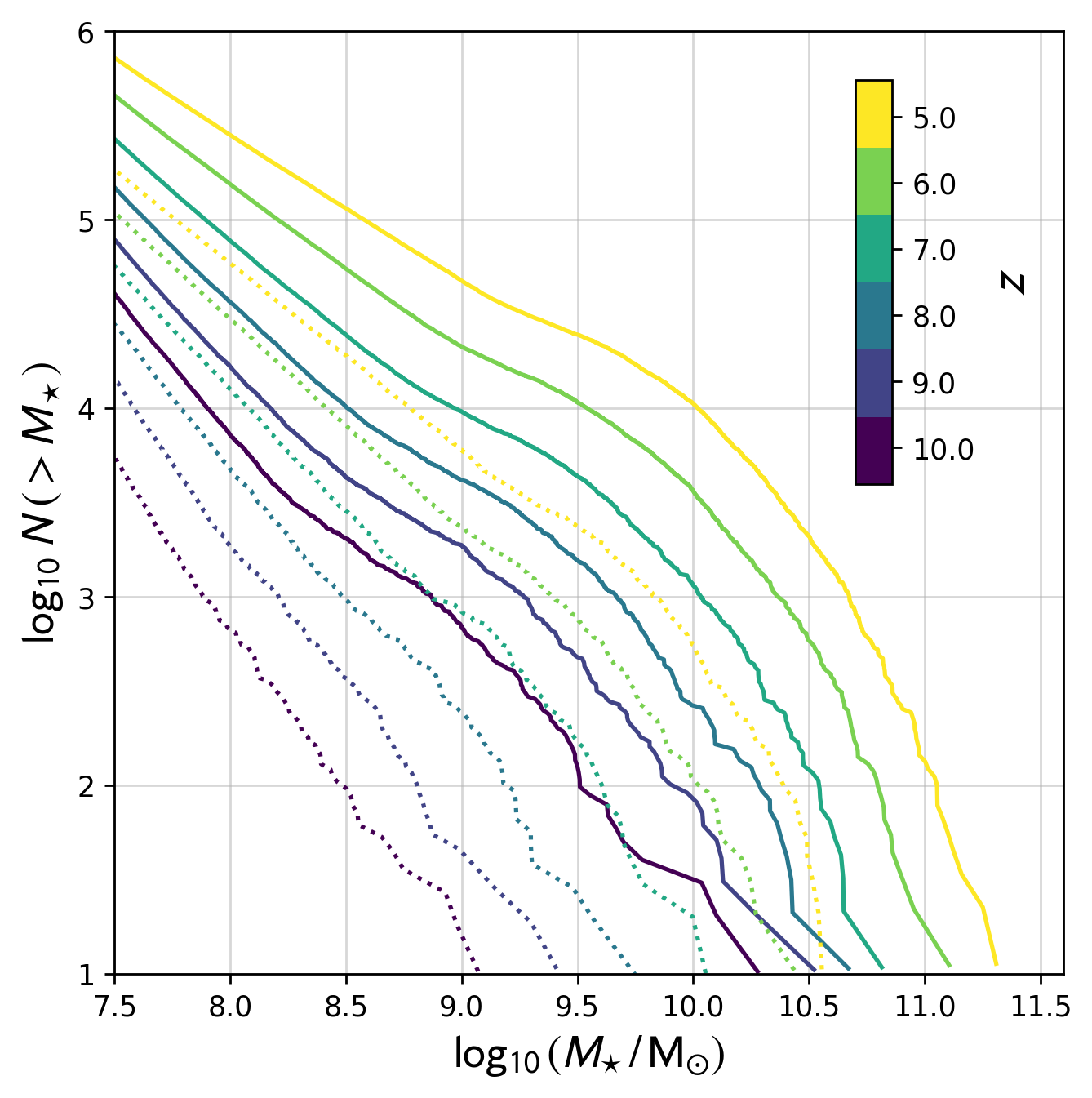}
    \caption{Cumulative distribution of stellar masses for all \flares\ regions combined (solid) and the fiducial \eagle\ Reference volume (dashed).}
    \label{fig:comp_hist}
\end{figure}

We begin by examining the raw number counts of galaxies.
\fig{comp_hist} shows the cumulative distribution function of galaxies with stellar mass for both \flares\ and the Reference periodic volume ($V = (100 \, \mathrm{cMpc})^3$).
We produce over $\sim 20$ times more $10^{10} \, \mathrm{M_{\odot}}$ galaxies at $z = 5$ than obtained in the 100\,cMpc periodic volume, despite the fact that the total high-resolution volume of all resimulated regions is only 50\% larger than the periodic volume.
This confirms that the first galaxies are significantly biased to higher overdensity regions.

\subsection{The Galaxy Stellar Mass Function}
\label{sec:results:gsmf}

The Galaxy Stellar Mass Function (GSMF) describes the number of galaxies per unit volume per unit stellar mass interval d$\log_{10}M$,
\begin{align}
    \phi(M) = N \,/\, \mathrm{Mpc^{-3} \, dex^{-1}}\;\;,
\end{align}
and is commonly described using a Schechter function \citep{schechter_analytic_1976},
\begin{align}
  \phi(M) \, \mathrm{d}\log_{10}M = \mathrm{ln}(10) \, \phi^* \, e^{-M/M^*} \, \left(\frac{M}{M^*} \right)^{\alpha + 1},
\end{align}
which describes the high- and low-mass behaviour with an exponential and a power law dependence on stellar mass, respectively.
Recent studies have found that a double Schechter function can better fit the full distribution \citep[\textit{e.g.} the GAMA survey, ][]{baldry_galaxy_2008}.
\begin{align}
    \phi(M) \, \mathrm{d}\log_{10}M& = \mathrm{ln}(10) \, e^{-M/M^*} \times \nonumber\\
    & \left[ \, \phi^{*}_{1} \, \left(\frac{M}{M^*} \right)^{\alpha_{1} + 1} + \phi^{*}_{2} \, \left( \frac{M}{M^*} \right)^{\alpha_{2} + 1} \right].
\end{align}
The low mass slope of the second schechter function contributes to only a very narrow dynamic range.
Above this range the exponential dominates, and below this the low mass slope of the first schechter function dominates.
It is therefore poorly constrained by the binned data, and so as not to introduce further degrees of freedom into our fit we fix it at $\alpha_2=-1$.
We define the stellar mass $M_\star$ as the total mass of all star particles, associated with the bound subhalo, within a 30 kpc aperture (proper) centred on the potential minimum of the subhalo.\footnote{Two substructures within 30 kpc of each other are still identified as separate structures, and only the particles associated with each structure contributes to its aperture-measured properties.}

\subsubsection{The cosmic GSMF}
\label{sec:cos_gsmf}

In this section, we present results for the universal GSMF, averaged within our (3.2\,cGpc)$^3$ box.  This is obtained by combining the individual GSMFs from each of our resimulation volumes with appropriate weighting, as described in \Sec{method:weighting}.

\begin{figure}
	\includegraphics[width=\columnwidth]{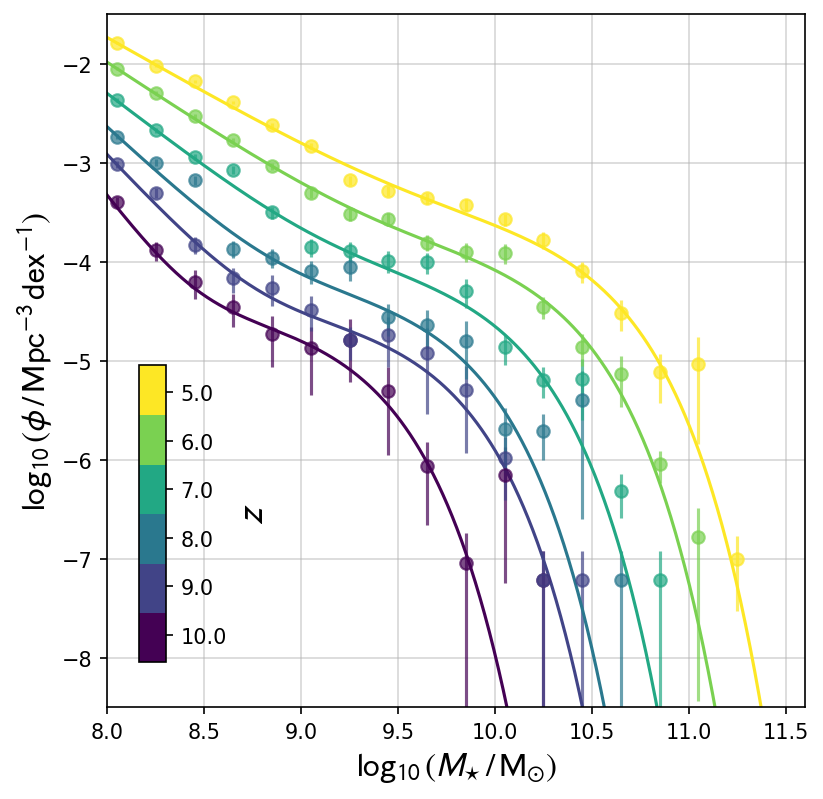}
  \includegraphics[width=\columnwidth]{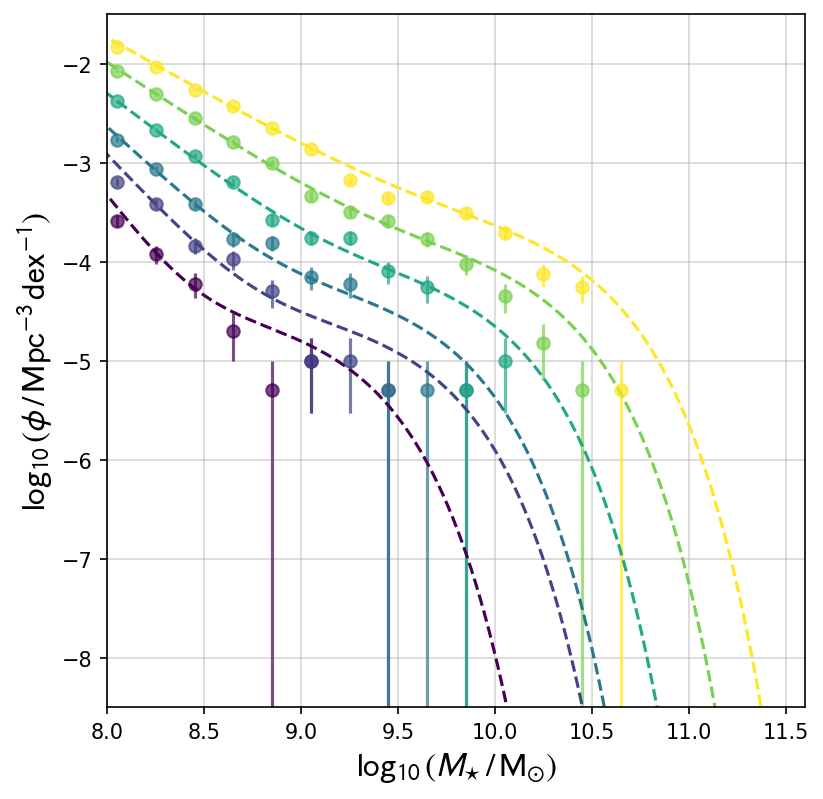}
    \caption{\textit{Top:} Redshift evolution of the \flares\ composite galaxy stellar mass function.
    Points show binned differential counts with Poisson $1\sigma$ uncertainties from the simulated number counts.
    Solid lines show double-Schechter function fits, quoted in \tab{schechter_params}. The parameter evolution is shown in \fig{fit_param_evolution}.
    \textit{Bottom:} as for the top panel, but points show the counts from the periodic Reference volume. The dashed lines show the double-Schechter fitted relation from \flares. The coverage of the massive end in the periodic volume is poor.
    }
    \label{fig:gsmf}
\end{figure}

\begin{figure}
	\includegraphics[width=\columnwidth]{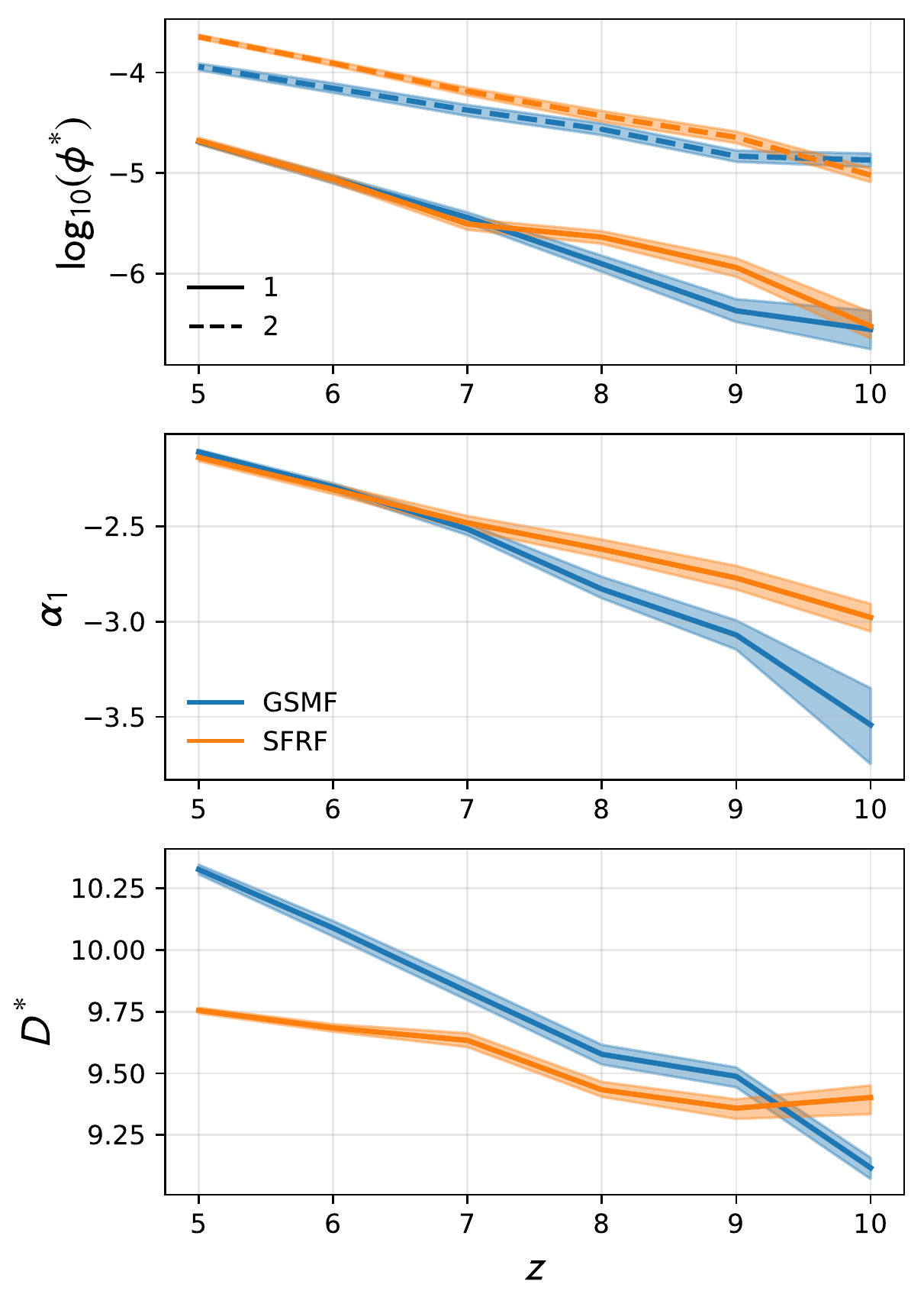}
    \caption{Parameter evolution for double-Schechter function fits to the \flares\ composite galaxy stellar mass function (GSMF, blue) and star formation rate function (SFRF, orange).
    The low (1) and high (2) mass components are shown with solid and dashed lines, respectively.
    Shaded regions show the $16^{\mathrm{th}}-84^{\mathrm{th}}$ percentile uncertainty obtained from the fit posteriors (see \App{fitting} for details)
    The low-mass slope of the high-mass component ($\alpha_{2}$) is fixed at -1.
    The characteristic mass of the GSMF ($\mathrm{M_\star}$) and the characteristic SFR of the SFRF ($\SFR_*$) are shown in the bottom panel, labelled $D_*$.
    $\SFR_*$ is offset by $+10^{8}$ to aid comparison with $M_\star$.
    The GSMF and SFRF show very similar behaviour; the normalisation of both components and the low-mass slope all increase with decreasing redshift.
    The characteristic mass increases with decreasing redshift for the GSMF, whereas the characteristic star formation rate of the SFRF shows a flatter redshift relation.
    }
    \label{fig:fit_param_evolution}
\end{figure}

The top panel of \fig{gsmf} shows the GSMF for redshifts between $z = 10 \mapsto 5$.
We show differential counts in bins $0.2 \, \mathrm{dex}$ in width (with 1$\sigma$ poisson uncertainties).
The solid lines show double-Schechter function fits at each integer redshift.
The normalisation increases with decreasing redshift, and the characteristic mass (or knee) of the distribution shifts to higher masses.
This is more clearly seen in \fig{fit_param_evolution}, which shows the evolution of the double-Schechter parameters with redshift.
The low-mass slope also gets shallower with decreasing redshift, from $-3.5$ at $z = 10$ to $-2.0$ at $z = 5$.


Our composite GSMF significantly extends the dynamic range of the GSMF compared to the periodic volumes.
To demonstrate, the bottom panel of \fig{gsmf} shows the \flares\ double-Schechter fits, alongside the binned counts from the Reference periodic volume.
At each redshift the maximum stellar mass probed is approximately an order of magnitude larger in \flares.
In fact, the periodic reference volume barely probes the exponential tail of the high mass component of the GSMF.
When fitting a double-Schechter to the binned Reference volume counts we found that the parameters of the high mass component were completely unconstrained.
However, it is clear from the bottom panel of \fig{gsmf} that the low-mass slope is consistent between the Reference volume and \flares.
We have also tested that this is the case for the $(50 \; \mathrm{Mpc})^{3}$ AGNdT9 periodic volume.
This provides evidence that our weighting method is accurately recovering the composite GSMF, without suffering from completeness bias.
We note that the GSMF in the AGNdT9 and Reference periodic volumes is also in agreement at the low mass end, which gives us confidence that model incompleteness is not affecting our results.

\begin{figure*}
	\includegraphics[width=\textwidth]{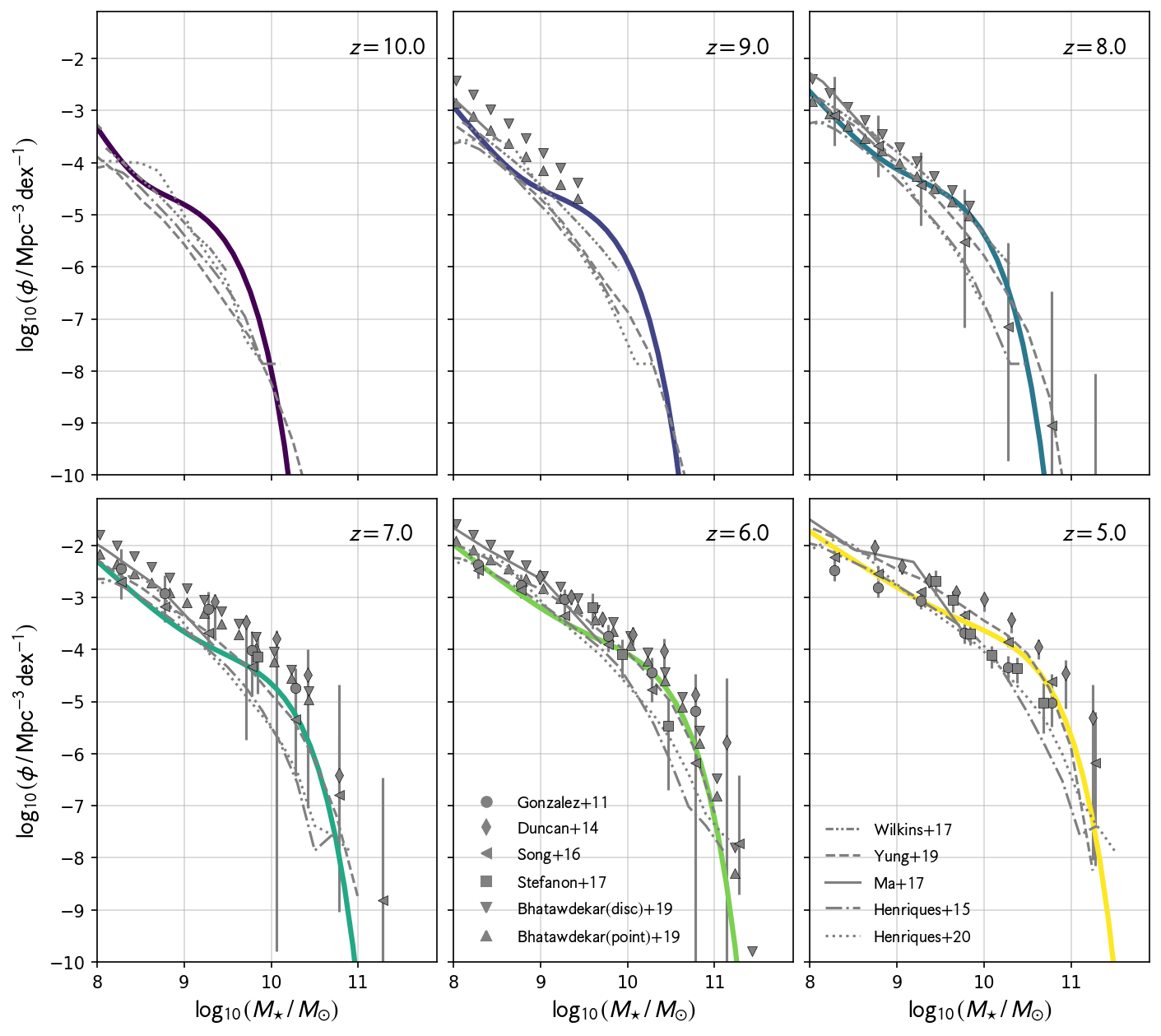}
    \caption{\flares\ composite galaxy stellar mass function evolution, alongside observational constraints \citep{gonzalez_evolution_2011,duncan_mass_2014,song_evolution_2016,stefanon_rest-frame_2017,bhatawdekar_evolution_2018} as well as predictions from other models \citep{wilkins_properties_2017,ma_simulating_2018,yung_semi-analytic_2019,henriques_galaxy_2015,henriques_l-galaxies_2020}.
    There is some disagreement over the normalisation of the GSMF between different observational studies, however \flares\ is consistent up to $z = 9$.
    }
    \label{fig:gsmf_multi_both}
\end{figure*}

In \fig{gsmf_multi_both} we show the composite \flares\ GSMF against a number of high-$z$ observational constraints in the literature \citep{gonzalez_evolution_2011,duncan_mass_2014,song_evolution_2016,stefanon_rest-frame_2017,bhatawdekar_evolution_2019}.
These studies show a spread of $\sim 0.5 \; \mathrm{dex}$ at $z = 5$, which highlights the difficulty of accurately measuring the GSMF at high redshift.
The \flares\ composite GSMF lies within this inter-study scatter, most closely following the relations derived by \cite{song_evolution_2016} up to $z = 7$.
At $z \geqslant 8$ observational constraints are limited to cluster lensing studies such as the Hubble Frontier Fields, which do not probe the high-mass end due to the limited volume probed, but can reach very lower stellar masses ($\sim 10^{7} \; M_{\odot}$).
The fits presented in \cite{bhatawdekar_evolution_2019} have a higher normalisation than in \flares\ over the accessible mass range, though they quote an uncertainty at $10^{8.5} \; M_{\odot}$ of $\sim 0.6$ dex at $z =9$; \flares\ lies within this uncertainty for the point sources, but is still in tension with the normalisation for disc-like sources.\footnote{We show both disc-like and point-like constraints on the \cite{bhatawdekar_evolution_2018} GSMF; we will present our galaxy sizes in future work, though we note here that many of our galaxies have disc-like morphologies even at the highest redshifts.}
There is good agreement with the low-mass slope for both sources.

We also compare in \fig{gsmf_multi_both} to predictions from other galaxy formation models.
The Feedback In Realistic Environments (\textsc{Fire}) project performed zoom simulations of individual halos with masses between $10^{8} \;-\; 10^{12} \, M_{\odot}$, which were then combined to provide a composite galaxy stellar mass function probing the low-mass regime \citep{ma_simulating_2018}.
\flares\ is consistent with \textsc{Fire} at all redshifts where their mass range overlaps.
\fig{gsmf_multi_both} also shows both the 2015 and 2020 versions of \lgals.
Both models are in reasonably good agreement at all redshifts shown, but tend to underestimate the number density of massive galaxies at $z=5$ compared to both \flares\ and the observations.

\cite{yung_semi-analytic_2019} presented results from the Santa Cruz semi-analytic model \citep{somerville_star_2015}, which extends to a wide dynamic range.
Whilst \flares\ is consistent with this model for $z \leqslant 7$, at $z \geqslant 8$ the Santa Cruz model predicts a more power-law shape to the GSMF, with a lower normalisation at the characteristic mass.
This is in agreement with the observed flattening of the GSMF with increasing redshift.

\subsubsection{Environmental dependence of the GSMF}
\label{sec:env_gsmf}

\begin{figure}
	\includegraphics[width=\columnwidth]{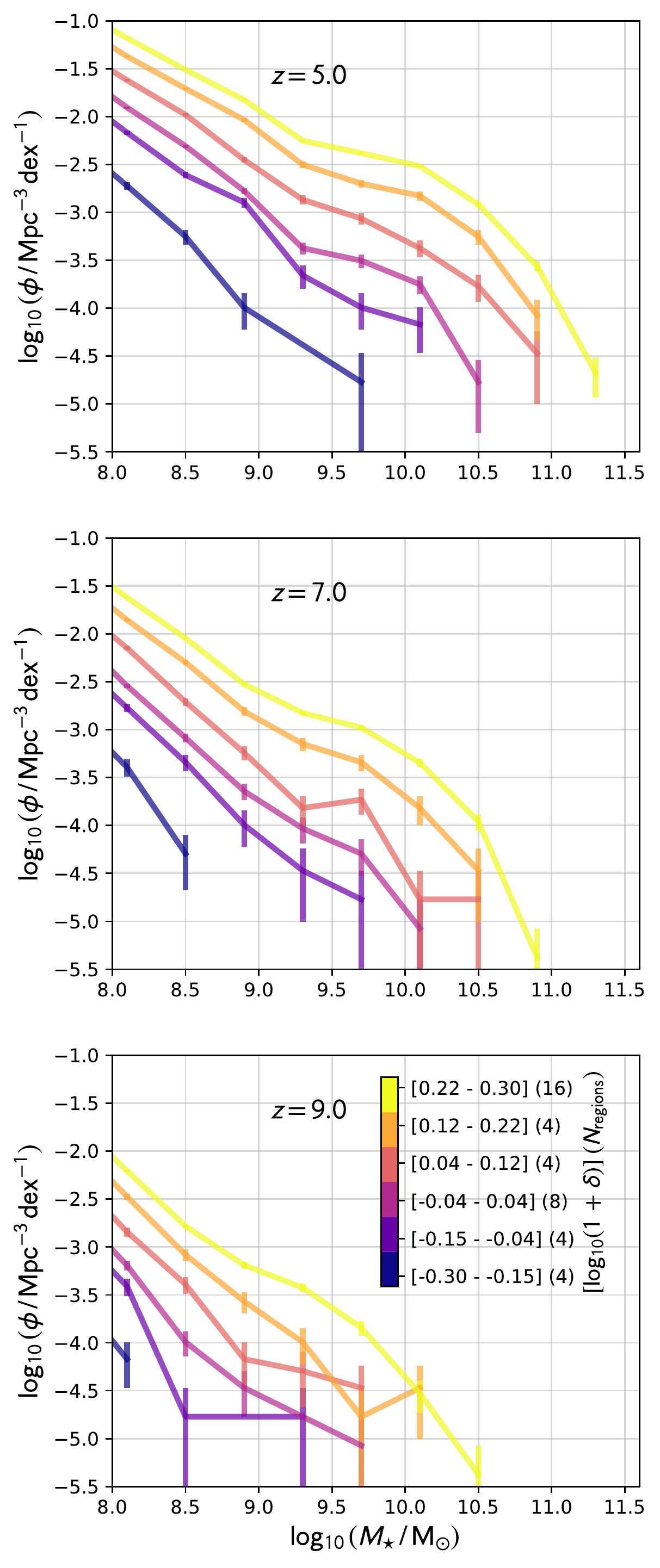}
    \caption{The \flares\ GSMF between $z = 5$ and $9$ split by binned log-overdensity.
    The binning is shown in the legend, along with the number of regions in each bin.
    Poisson 1$\sigma$ uncertainties are shown for each bin from the simulated number counts.
    The normalisation increases with increasing overdensity, and probes higher stellar masses.}
    \label{fig:gsmf_overdensity}
\end{figure}

Our zoom simulations of a range of overdensities not only allow us to construct a composite GSMF for the entire $(3.2 \, \mathrm{Gpc})^{3}$ volume, but also investigate the environmental effect on the GSMF.
\sec{method} demonstrates the wide range of environments probed, from extremely underdense void regions, to the most overdense high redshift structures that are likely to collapse in to massive, $> 10^{15} \; \mathrm{M_{\odot}}$ clusters by $z = 0$ \citep{chiang_ancient_2013,lovell_characterising_2018}.

\fig{gsmf_overdensity} shows the GSMF in bins of log-overdensity from $z = 5 - 9$.
We use wider bins than previously ($0.4 \, \mathrm{dex}$) due to the lower galaxy numbers in each resimulation.
As expected, higher overdensity regions have a higher normalisation, $\sim +2 \; \mathrm{dex}$ above the lowest overdensity regions at $M_\star \,/ \mathrm{M_{\odot}} = 10^{9.5} \; (z = 5)$.
There is also an apparent difference in the shape as a function of log-overdensity:
lower overdensity regions exhibit a distribution that is more power-law -like, whereas higher overdensity regions clearly show a double-Schechter -like knee.
This may be due to the higher number of galaxies in the overdense regions, better sampling the knee, but may also point to differing assembly histories for galaxies in different environments.
We will explore the star formation and assembly histories more closely in future work.

The dependence of the GSMF on overdensity may explain the tension between the composite \flares\ GSMF and other models at $z > 7$ seen in \fig{gsmf_multi_both}.
Our much larger box allows us to sample extreme overdensities that are not present in smaller volumes.
Observationally, the \cite{song_evolution_2016} results show a more power law-like form at $z = 8$.
Double-Schechter forms of the GSMF at low-$z$ have been attributed to the contribution of a passive and star forming population, each fit individually by a single Schechter function \citep{kelvin_galaxy_2014,moffett_galaxy_2016}, though this separation is not perfect \citep[\textit{e.g.}][]{ilbert_mass_2013,tomczak_sfr-m*_2016}.
The robust double-Schechter shape measured in \flares\ at $z \geqslant 8$ is therefore curious; we see in \app{ssfr_cut} that there is no significant passive population as a function of stellar mass.
We therefore tentatively suggest that the tension may be due to the small volume probed observationally at these depths, which does not probe extreme environments that contribute significantly to the cosmic GSMF.

We do not fit each binned GSMF in log-overdensity as there are insufficient galaxies to provide a robust fit.
However, we do provide fits to the normalisation at a given stellar mass and redshift, in the following form,
\begin{equation}
  \mathrm{log_{10}} \; \phi \,(\mathrm{log_{10}}(1+\delta) \,|\, M_\star, z) = m \,[\mathrm{log_{10}}(1+\delta)] + c,
\end{equation}
where $\mathrm{log_{10}}(1\,+\,\delta)$ is the overdensity of the region.
\tab{norm_GSMF} shows these fits for bins $\pm 0.2$ dex wide centred at $\mathrm{log_{10}}(M_\star \,/\, \mathrm{M_{\odot}}) = [8.5,\,9.7]$.

\begin{table}
	\centering
	\caption{Fits to the normalisation, $\log_{10}(\phi/$Mpc$^{-3}$dex$^{-1}$), of the GSMF at different redshifts and masses (see \sec{env_gsmf}).}
	\label{tab:norm_GSMF}
	\begin{tabular}{cccc} 
		\hline
		$z$ & $\mathrm{log_{10}(M_\star / M_{\odot})}$ & $m$ & $c$ \\
		\hline
        5 & 8.5 & 3.5 & -2.4 \\
        7 & 8.5 & 4.4 & -3.2 \\
        9 & 8.5 & 4.6 & -4.0 \\
        5 & 9.7 & 4.8 & -3.6 \\
        7 & 9.7 & 4.4 & -4.2 \\
        9 & 9.7 & 4.0 & -4.9 \\
    \hline
	\end{tabular}
\end{table}

\subsection{The Star Formation Rate Distribution Function}
\label{sec:cos_sfrf}

\begin{figure}
	\includegraphics[width=\columnwidth]{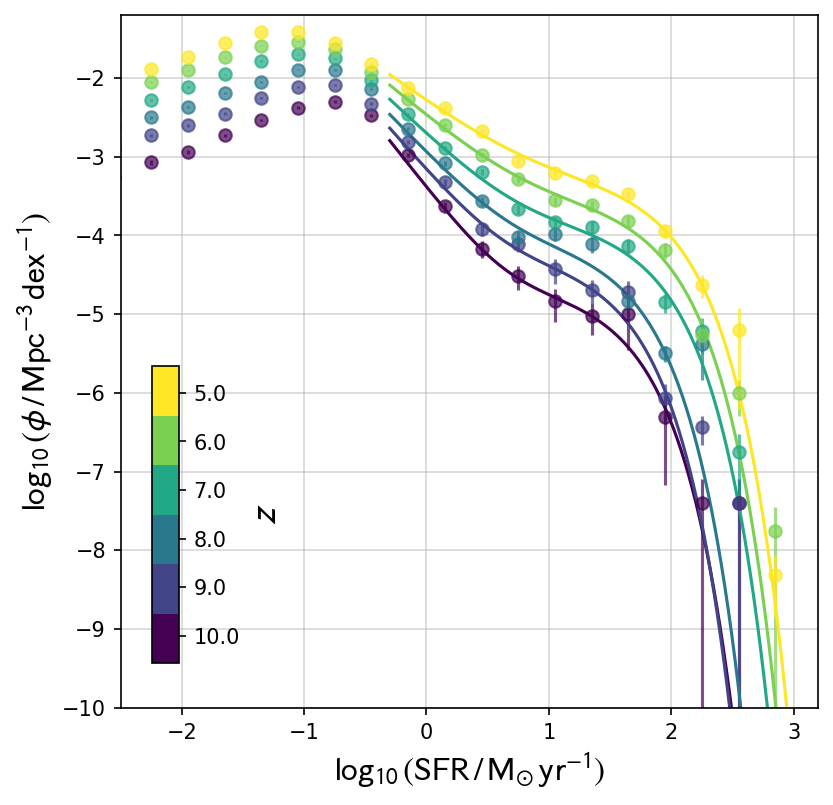}
    \caption{Redshift evolution of the \flares\ composite star formation rate distribution function.
		Points show binned differential counts with Poisson 1$\sigma$ uncertainties from the simulated number counts.
		Solid lines show double-Schechter function fits, quoted in \tab{sfrf_schechter_params}.}
    \label{fig:sfrf_all}
\end{figure}

The Star Formation Rate distribution Function (SFRF) describes the number of galaxies per unit volume per unit star formation rate interval d$\log_{10}\SFR$, where $\SFR$ is the star formation rate,
\begin{align}
     \phi(\SFR) = N \,/\, \mathrm{Mpc^{-3} \, dex^{-1}}\;\;.
\end{align}
We define the SFR as the sum of the instantaneous SFR of all star forming gas particles, associated with the bound subhalo, within a 30 kpc aperture (proper) centred on the potential minimum of the subhalo.

\subsubsection{The cosmic SFRF}
\label{sec:results:sfrf:cosmic}

In \fig{sfrf_all} we plot the evolution of the \flares\ composite SFRF.
We provide counts in bins 0.3 dex in width.
There is a clear low-mass turnover between $\sim 0.1\,-\,0.3 \, \mathrm{M_{\odot} \; yr^{-1}}$, but above this the shape is well described by a double-Schechter function.
\cite{salim_star_2012} argue that a single-Schecter is inadequate to describe the SFRF, as we find, though they propose a 'Saunders' function that does not provide a good fit to the \flares\ SFRF.
We provide fits using the following parametrisation,
\begin{align}
    \phi(\SFR) \, \mathrm{d}\log_{10}&\SFR = \mathrm{ln}(10) \, e^{-\SFR/\SFR^*} \times \nonumber\\
    & \left[ \, \phi^{*}_{1} \, \left(\frac{\SFR}{\SFR^*} \right)^{\alpha_{1} + 1} + \phi^{*}_{2} \, \left(\frac{\SFR}{\SFR^*} \right)^{\alpha_{2} + 1} \right].
\end{align}
We limit our fits to those galaxies with $\SFR > 0.5 \, \mathrm{M_{\odot} \; yr^{-1}}$; these fits are provided in \tab{sfrf_schechter_params}.
We also plot the parameter evolution with redshift in \Fig{fit_param_evolution}.
The characteristic star formation rate, $\SFR_*$, is offset by $+10^{8}$ to aid comparison with the GSMF characteristic mass, $\mathrm{M_\star}$.

The normalisation of both components ($\phi_{1};\,\phi_{2}$), as well as the low-SFR slope ($\alpha_{1}$), increase with decreasing redshift.
These trends are surprisingly similar to those seen for the equivalent parameters in the GSMF.
The low-SFR normalisation is almost identical, as is the high-SFR normalisation, with a small $\sim +0.2\;\mathrm{dex}$ offset.
The low-SFR slope $\alpha_1$ is shallower than that of the GSMF at the highest redshifts ($z\geqslant 8$), but identical at lower redshifts.
However, the evolution of the characteristic SFR is significantly flatter compared to that of the characteristic mass for the GSMF.
This suggests a redshift-independent upper limit to the SFR.
The strong correspondence between the shape of the GSMF and the SFRF may be the result of the tight star-forming sequence relation at all redshifts (see \sec{results:sfs}).

\begin{figure*}
	\includegraphics[width=\textwidth]{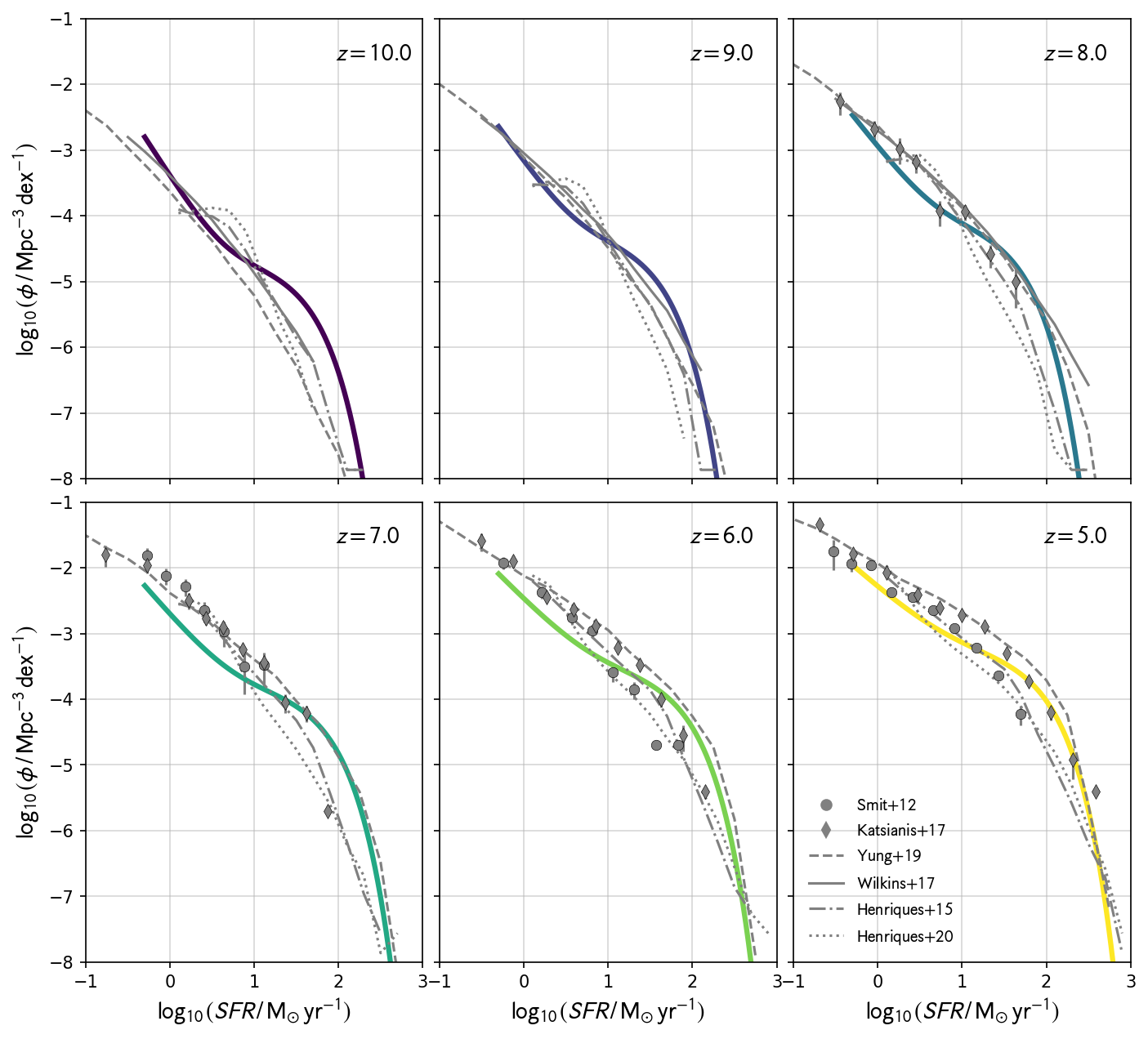}
        \caption{Evolution of the \flares\ composite star formation rate distribution function (coloured, solid lines), compared with observational constraints from UV data and other model predictions.
				\protect\cite{smit_star_2012} derive SFRs from UVLF data, as do \protect\cite{katsianis_evolution_2017} using \protect\cite{bouwens_uv_2015} data.
				Both are corrected to a Chabrier IMF using the conversion factors quoted in \protect\cite{kennicutt_jr_star_2012}.
				The Santa-Cruz SAM \protect\citep[][dashed line]{yung_semi-analytic_2019} and \bluetides\ simulation \protect\citep{wilkins_properties_2017} show a different behaviour, with a power law shape at higher redshifts, in contrast to the prominent knee seen in \flares\ up to $z = 10$.
				Both \lgals\ models also show similar behaviour, though with lower normalisation at the high-SFR end \citep{henriques_galaxy_2015,henriques_l-galaxies_2020}.
				}
        \label{fig:sfrf_multi_both}
\end{figure*}

This double-Schechter form of the SFRF is in some tension with observational constraints.
\fig{sfrf_multi_both} shows a comparison with UV derived relations from \cite{smit_star_2012} and \cite{katsianis_evolution_2017} (the latter using \citealt{bouwens_uv_2015} data).
For low-SFRs the observed normalisation is slightly higher ($\sim \, 0.3$ dex) from $z = 5$ to $7$.
There is no prominent knee in the observed relations, and the exponential tail drops off at lower SFRs than in the simulations.

\fig{sfrf_multi_both} also shows results from recent cosmological models.
As with the GSMF, there is some tension with the SFRF produced by the Santa Cruz models \citep{yung_semi-analytic_2019}.
\flares\ has a distinct double-Schechter shape, whereas the SC model appears as a single schechter at $z = 5$, before evolving to a power law at $z = 10$.
The \bluetides\ results \citep{wilkins_properties_2017} also show a similar power law relation at $z \geqslant 8$, in tension with the prominent knee in \flares.
Both \lgals\ models show similar power law-like behaviour, though with lower normalisation at the high-SFR end \citep{henriques_galaxy_2015,henriques_l-galaxies_2020}, though in better agreement with the existing observational data at $z = 6$ compared to the Santa Cruz model and \flares.

The offset in normalisation of the \flares\ SFRF at high SFRs with the observations may be a selection effect due to highly dust-obscured galaxies.
These galaxies, with number densities of $\sim 10^{-5} \; \mathrm{cMpc^{-3}}$  at $z \sim 2$ \citep{simpson_alma_2014}, will be missed in higher redshift rest frame-UV observations.
We will perform a direct comparison with the UV luminosity function, including selfconsistent modelling of dust attentuation, in Paper II, Vijayan et al., \textit{in prep.}.
The offset may also be a modelling issue; \eagle\ was not compared to high redshift observables during calibration, only to data at much lower redshifts ($z = 0.1$) than those studied here ($z \geqslant 5$).
Improvements to the subgrid modelling at high-redshift, particularly that of star-formation feedback, may improve the agreement.

To investigate what effect our sampling of highly overdense regions has on the composite shape of the SFRF, we now look at the overdensity dependence of the SFRF.

\subsubsection{Environmental dependence of the SFRF}
\label{sec:env_sfrf}

\begin{figure}
	\includegraphics[width=\columnwidth]{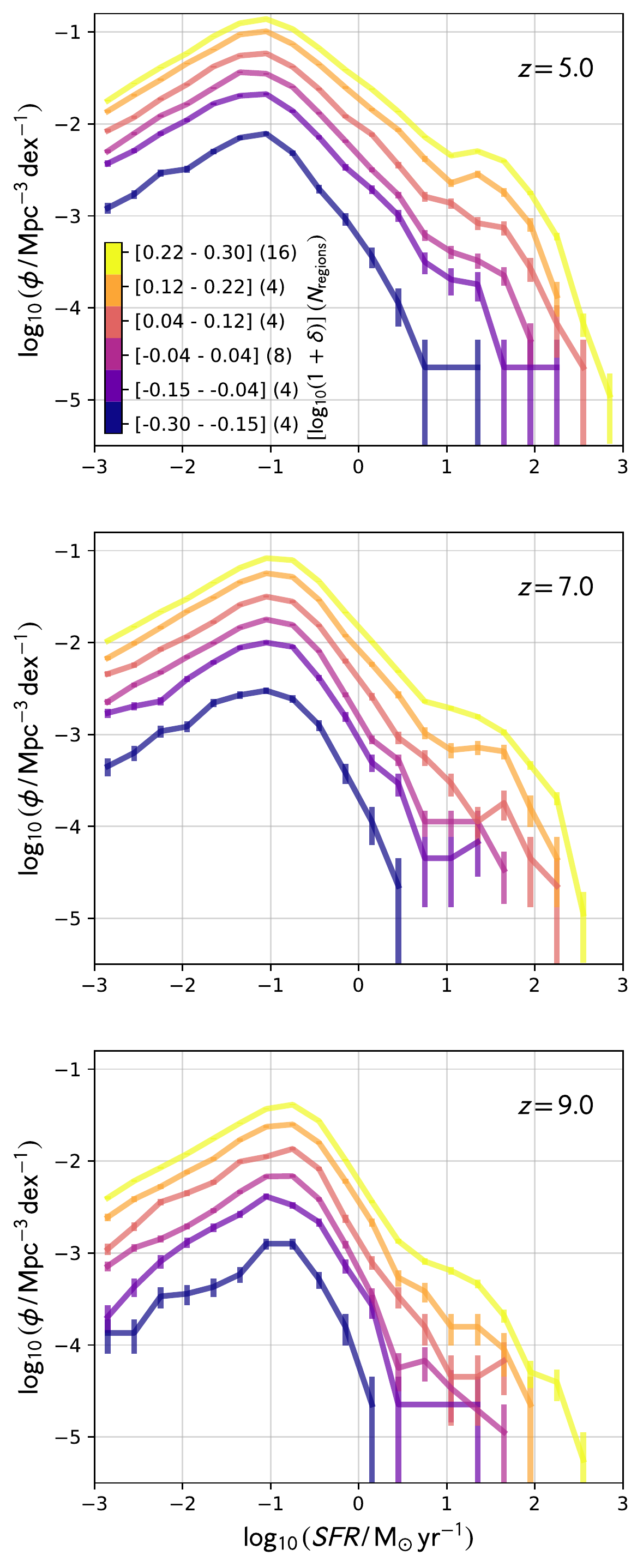}
    \caption{The \flares\ SFRF between $z = 5$ and $9$ split by binned log-overdensity.
		The binning is shown in the legend, along with the number of regions in each bin.
		Poisson 1$\sigma$ uncertainties are shown for each bin from the simulated number counts.
		The normalisation increases with increasing overdensity, and the maximum SFR increases.}
    \label{fig:sfrf_overdensity}
\end{figure}

\fig{sfrf_overdensity} shows the SFRF for regions binned by their log-overdensity.
There is almost no variation in the shape as a function of overdensity except for the highest overdensities, which show a more prominent double-Schechter knee in the high-SFR regime.
This behaviour is identical to that seen for the GSMF.
This may explain why the shape of the \flares\ composite SFRF differs with those of other cosmological models.
\flares\ better samples the rare, high-density regions that contribute significantly to the high-SFR ($\SFR > 100 \, \mathrm{M_{\odot} yr^{-1}}$) tail of the SFRF.
Both \bluetides\ and the Santa-Cruz model are run on regions with much smaller volumes ($500^{3}$ and $357^{3}$ $\mathrm{cMpc^3}$, respectively), which may not probe the extreme regions sampled in the \flares\ parent volume.
The mean density region in \fig{sfrf_overdensity} appears power law-like at all redshifts, which may present a better comparison with these models.

As for the GSMF, we provide fits to the normalisation at a given SFR and redshift, in the following form,
\begin{equation}
  \phi \,(\mathrm{log_{10}}(1+\delta) \,|\, \SFR, z) = m \,[\mathrm{log_{10}}(1+\delta)] + c,
\end{equation}
where $\mathrm{log_{10}}(1\,+\,\delta)$ is the overdensity of the region.
\tab{norm_SFRF} shows these fits for bins $\pm 0.2$ \; dex wide centred at $\mathrm{log_{10}}(\SFR\,/\,\mathrm{M_{\odot}} \, yr^{-1}) = [-0.5,\,0.5]$.
The normalisation increases with increasing overdensity as expected.
The trends with redshift are also broadly similar to those seen for the GSMF.
\footnote{The only exception being the gradient of the GSMF relation at $M_\star \,/\, \mathrm{M_{\odot}} = 10^{9.7}$, which decreases with redshift,  whereas the redshift dependence is positive for the SFRF at all SFRs.}

\begin{table}
	\centering
	\caption{Fits to the normalisation, $\log_{10}(\phi_\SFR/\mathrm{Mpc^{-3} \, dex^{-1}}$) of the SFRF at different redshifts and star formation rates (see \protect\sec{env_sfrf}).}
	\label{tab:norm_SFRF}
	\begin{tabular}{lccc} 
		\hline
		$z$ & $\log_{10}(\SFR \,/\, \Msun \, \mathrm{yr}^{-1})$ & $m$ & $c$ \\
		\hline
        5 & -0.5 & 3.0 & -2.0 \\
        7 & -0.5 & 3.2 & -2.2 \\
        9 & -0.5 & 3.5 & -2.6 \\
        5 & 0.5 & 3.8 & -2.8 \\
        7 & 0.5 & 4.4 & -3.4 \\
        9 & 0.5 & 4.5 & -4.0 \\
        \hline
	\end{tabular}
\end{table}

\subsection{The Star-Forming Sequence}
\label{sec:results:sfs}

Observations at both high- and low-$z$ suggest a tight relation between star formation rate and stellar mass, known as the `main sequence', or star-forming sequence \citep[SFS,][]{brinchmann_physical_2004,noeske_star_2007,speagle_highly_2014}.
The SFS is typically parametrised as a linear relation,
\begin{align}
  \mathrm{log_{10}(\SFR)} = \alpha \; \mathrm{log_{10}}(M_\star\,/\, \mathrm{M_{\odot}}) + \beta\;\;.
\end{align}
Observations suggest that the normalisation $\beta$ increases with redshift, whilst the slope $\alpha$ remains relatively constant \citep{daddi_multiwavelength_2007, santini_star_2009, salmon_relation_2015}.

There have been suggestions of a turnover in the SFS at high stellar masses, though the turnover mass, and its evolution with redshift, are less clear \protect\citep{lee_turnover_2015,tasca_evolving_2015,santini_star_2017}.
Such a turnover is necessary to explain the GSMF at low redshift; a single power law slope would lead to too many massive galaxies being formed \citep[between $10^{10} < M_\star\,/\,\mathrm{M_{\odot}} < 10^{11}$,][]{leja_reconciling_2015}.
The turnover may be evidence for a change in the dominant channel of stellar mass growth, from smooth gas accretion to merger-driven growth.

\begin{figure}
	\includegraphics[width=\columnwidth]{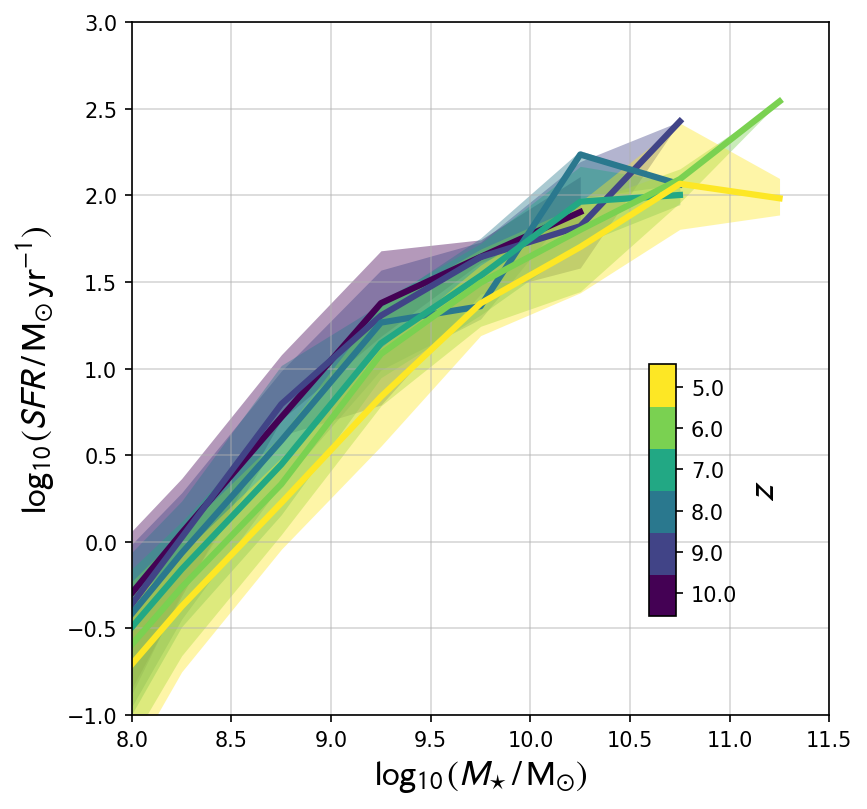}
  \includegraphics[width=\columnwidth]{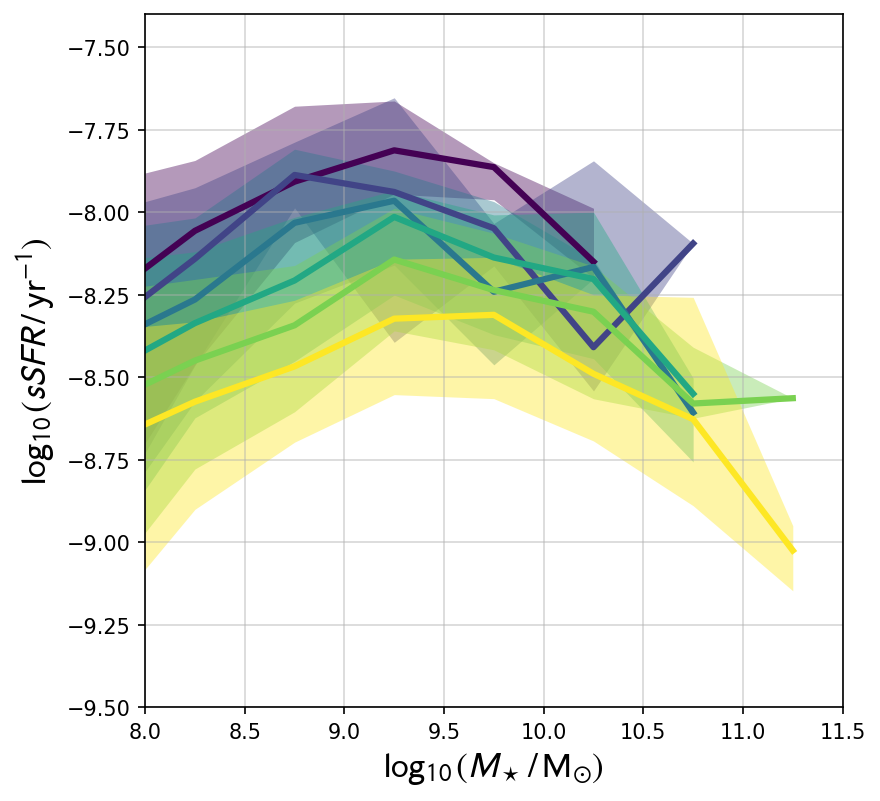}
    \caption{\textit{Top:} Redshift evolution of the \flares\ composite star forming sequence.
    Solid lines show the weighted composite SFS for centrals + satellites, with the 16$^{\mathrm{th}}$-84$^{\mathrm{th}}$ spread shaded.
    \textit{Bottom:} as for the top panel, but showing the specific-star formation rate - stellar mass relation.
    }
    \label{fig:sfs_all}
\end{figure}

The top panel of \fig{sfs_all} shows the redshift evolution of the SFS in \flares.
In the bottom panel of \fig{sfs_all} we also show the specific-star formation rate (sSFR) against $M_\star$ relation.
To construct the median lines, we weight each galaxy in the sample by the appropriate factor for the overdensity of the resimulation volume, as described in \Sec{method:weighting}.\footnote{In fact, as shown in \Fig{sfs_overdensity}, the environmental dependence is very weak and so the weighted relations are very similar to the unweighted ones.}
There is a clear trend of decreasing normalisation with decreasing redshift, approximately $0.5$ dex between $z = 10 \,-\, 5$.
There is some noise in the weighted relation at $z = 8$ for galaxies with $M_\star > 10^{9.5} \, M_{\odot}$; we checked, and found that this is due to a small number of galaxies in mean density regions above this mass limit with low SFRs, biasing the normalisation down.

We have not excluded `passive' galaxies from our measurement of the SFS.
We present results for the SFS assuming different specific-SFR cuts in \app{ssfr_cut}, though note here that they make negligible difference to the relations at $z \geqslant 5$ for even the most liberal cuts.

\begin{figure}
	\includegraphics[width=\columnwidth]{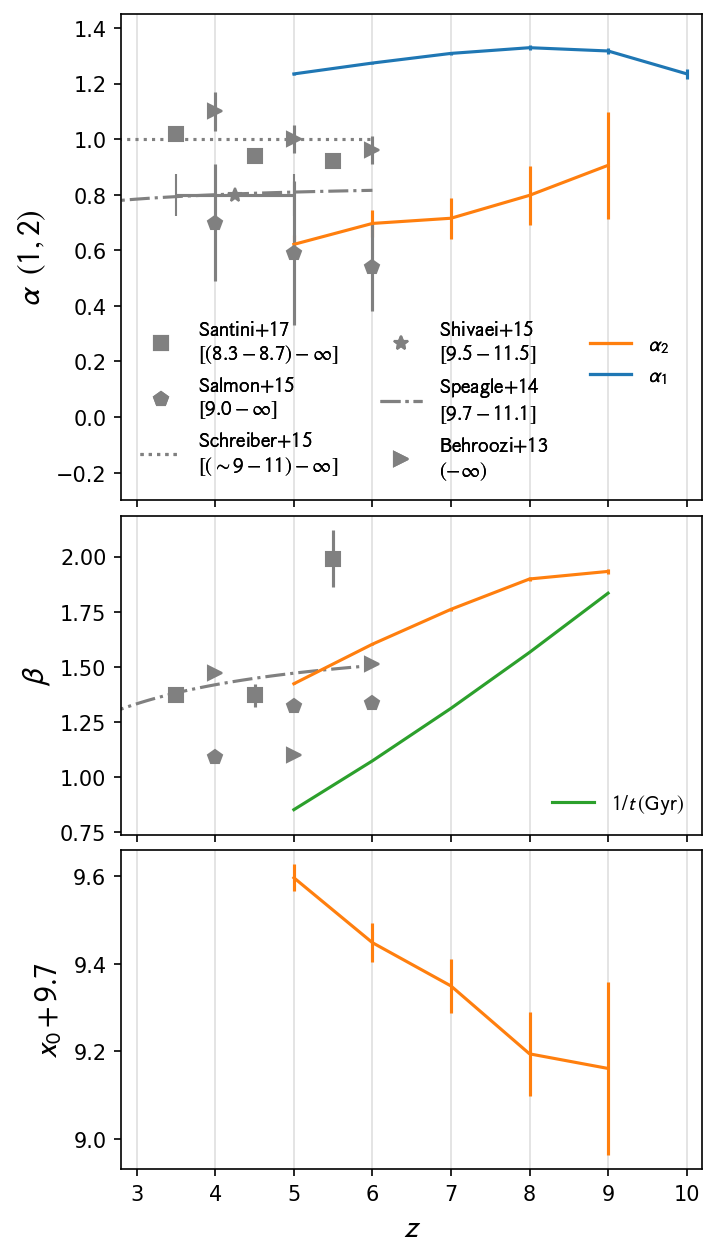}
    \caption{Redshift evolution of the piecewise-linear fit to the SFS.
    Observational results are plotted where available in grey, from \protect\cite{behroozi_average_2013,speagle_highly_2014,shivaei_mosdef_2015,salmon_relation_2015,schreiber_herschel_2015,santini_star_2017}.
    The lower- and upper-mass completeness limits for these studies are quoted in the legend.
    \textit{Top}: high- and low-mass slope, in orange and blue respectively.
    \textit{Middle}: normalisation, $\beta$, in orange.
    The inverse age of the Universe in Gyr is shown in green; the normalisation approximately follows the same relation, but with a slightly shallower evolution.
    \textit{Bottom}: turnover mass in log-solar masses, in orange.
    }
    \label{fig:sfs_fit_evolution}
\end{figure}

There is a clear turnover in the \flares\ star-forming sequence at high masses ($\sim \,>\, 10^{9.3} \mathrm{M_{\odot}}$).
We account for this by fitting a piecewise-linear relation, with an upper- and lower-mass part, for stellar mass re-normalised at $10^{9.7} \mathrm{M_{\odot}}$,
\begin{align}
	\log_{10}\SFR = \alpha_{1} \, \mathrm{log_{10}}(M_\star \,/\,10^{9.7} \mathrm{M_{\odot}}) + \beta_{1} \quad & x \leqslant x_{0} \\
	\log_{10}\SFR = \alpha_{2} \, \mathrm{log_{10}}(M_\star \,/\,10^{9.7} \mathrm{M_{\odot}}) + \beta_{2} \quad & x \geq x_{0} \;,
\end{align}
where $\alpha_{1}$ is the low-mass slope, $\alpha_{2}$ is the high-mass slope, and $x_{0}$ is the turnover mass in log-solar masses. The normalisation at the turnover, $\beta_{0}$, is then given by
\begin{align}
\beta_{0} & = \beta_{2} + \alpha_{2} \, x_{0} \\
& = \beta_{1} + \alpha_{1} \, x_{0} \;\;.
\end{align}

We use the \textsc{scipy} implementation of non-linear least squares to perform the fit, combined with a non-parametric bootstrap approach for estimating parameter uncertainties.
The bootstrap is implemented as follows: we select, with replacement, 10\,000 times from the original data, each resample being the same size as the original data.
We then fit each sample independently; parameter estimates are given by the median of the resampled fit distributions, and uncertainties are given as the 1$\sigma$ spread in the distributions (unless otherwise stated).
The parameter fits are quoted in \protect\tab{sfs_params}.

\begin{figure}
	\includegraphics[width=\columnwidth]{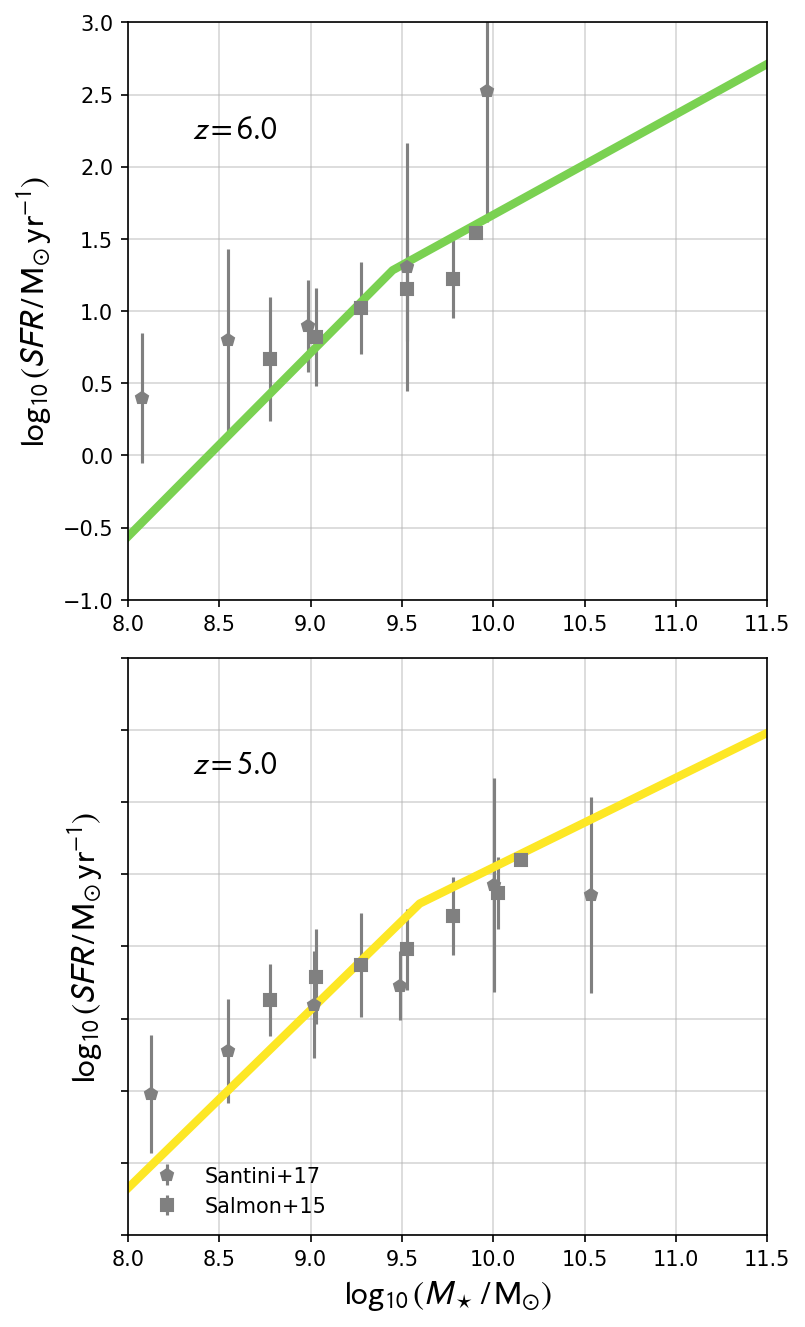}
    \caption{\flares\ composite SFS at $z = 5.0$ and $6.0$ compared to high redshift observational constraints from \protect\cite{santini_star_2017} and \protect\cite{song_evolution_2016}.
    }
    \label{fig:sfs_obs}
\end{figure}

\protect\fig{sfs_fit_evolution} shows the redshift evolution of each parameter against observational constraints where available.\footnote{The high mass slope and turnover are poorly constrained at $z = 10$ so we omit them.}
There are few robust observational constraints at $z > 6$, so we show constraints down to $z = 3$ to provide context to the redshift evolution \citep{behroozi_average_2013,schreiber_herschel_2015,shivaei_mosdef_2015,salmon_relation_2015,santini_star_2017}, including the compilation of pre-2014 measurements from \cite{speagle_highly_2014}.
These all represent single power-law measurements.
For all observations we quote the approximate lower mass completeness limit for the whole fit in the legend.
We also show a direct comparison of the fits to binned data from \cite{santini_star_2017} and \cite{salmon_relation_2015} in \fig{sfs_obs} at $z = 5\,-\,6$.

The normalisation is within the errors of the binned observations at these redshifts.
The fitted normalisation $\beta$ is also within the spread of the fitted relations at these redshifts, and continues the apparent increasing normalisation with increasing redshift from $z = 3$.
We also show the inverse age of the Universe (in Gyr); the fall in SFS normalisation approximately follows the same relation, but slightly shallower.

The slope of the observed relations shows considerable scatter spanning the range $\sim \, 0.5 - 1.1$.
We suggest that this is due to the lower-mass limit of these observations (quoted in the legend of \fig{sfs_fit_evolution}).
Since these studies fit a single power-law, and assume a high lower-mass completeness limit, ($M_\star \,/\, M_{\odot} > 10^{9.5}$), the measured slope will be biased to shallower slopes.
This can also be seen clearly in the binned relations in \fig{sfs_obs}; both \cite{santini_star_2017} and \cite{song_evolution_2016} straddle the turnover mass in \flares.
Finally, this can also be seen in the redshift evolution of these studies.
The observed slopes of \cite{salmon_relation_2015}, \cite{behroozi_average_2013} and \cite{santini_star_2017} all show a negative correlation with redshift.
The lower-mass completeness limit of these studies also increases with increasing redshift; as it increases, they tend to probe just the high-mass end of the SFS, rather than the steeper low-mass end.
This suggests that many high redshift measures of the SFS, where the mass completeness does not extend to very low masses, are only probing the SFS at stellar masses above the turnover, and the measured slopes do not represent a universal relation for all masses.

The turnover mass shows a negative correlation with redshift, increasing from $\sim 10^{9.2}$ to $10^{9.6} \, M_\star \,/\, M_{\odot}$ between $z = 9 - 5$.
\cite{ceverino_firstlight_2018} show no turnover in their FirstLight simulation results, but they do not probe above $10^{9.5}$ at $z = 6$, which is consistent with where we constrain the turnover.
There are unfortunately no observational constraints on the turnover mass at $z > 3$.
We note that the turnover mass is much lower than that measured in low-$z$ studies  \citep[$> 10^{10} \, M_\star \,/\, M_{\odot}$ at $z \leqslant 3$,][]{whitaker_constraining_2014,tasca_evolving_2015}.

\subsubsection{Environmental dependence of the SFS}

\begin{figure}
	\includegraphics[width=\columnwidth]{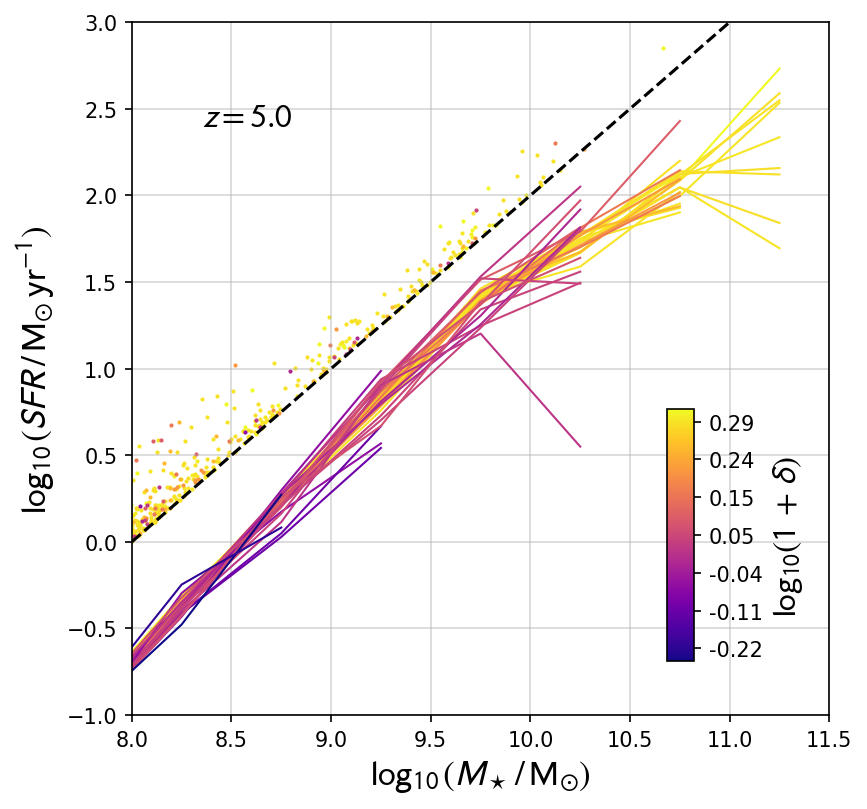}
    \caption{Coloured lines show the SFS for each region at $z = 5$, coloured by overdensity.
    The black dashed line shows a fixed $\mathrm{SSFR} = 10^{-8} \; \mathrm{yr^{-1}}$, and points above this show starbursting galaxies, coloured by their host region overdensity.
    }
    \label{fig:sfs_overdensity}
\end{figure}

\fig{sfs_overdensity} shows the SFS for each region individually at $z = 5$, coloured by overdensity.
The highest overdensities reach to higher stellar masses, as expected.
However, there is no dependence on overdensity of either the normalisation nor shape of the SFS, and we see this up to $z = 10$.
Observationally, at $z \sim 2$ there is a similar lack of dependence on environment as measured between protocluster and field regions \citep{koyama_evolution_2013,koyama_environmental_2014,shimakawa_mahalo_2017,shimakawa_mahalo_2018}, though these authors do note some differences in dense subgroups in protocluster candidates (we leave an investigation of the small-scale overdensity dependence of the SFS to future work).
However, at $z > 5$ \cite{harikane_silverrush._2019} find a $5\,\times$ enhancement in the SFR (at fixed stellar mass) of Lyman-$\alpha$ emitters in protoclusters compared to the field, though they only probe the low-mass regime ($M_{*} < 10^{9} \; M_{\odot}$).
It is as yet unclear whether these galaxies represent the main star-forming sequence, or starbursts that lie above it.
\cite{harikane_silverrush._2019} show that dusty star forming galaxies traced in the sub-mm are also spatially correlated with these structures \citep{geach_scuba-2_2017}, and lead to significant enhancements in the cosmic star formation rate density compared to the \cite{madau_cosmic_2014} relation.
In \flares, whilst the normalisation of the SFS at $z = 5 - 6$ is low in the stellar mass regime $M_{*} < 10^{9} \; M_{\odot}$ compared to observational constraints \citep{song_evolution_2016,santini_star_2017}, \fig{sfs_overdensity} shows that \flares\ does produce a number of galaxies with SFRs at least $5\times$ higher than on the main relation, and these are biased to high density regions.

We leave a thorough exploration of the passive and star-bursting galaxy populations in \flares\ to future work.

\section{Conclusions}
We have presented the first results from the \flares\ simulations, resimulations with full hydrodynamics of a range of overdensities during the Epoch of Reionisation (EoR, $z \geqslant 5$) using the \eagle\ \citep{schaye_eagle_2015} physics.
We described our novel weighting procedure that allows the construction of composite distribution functions that mimic extremely large periodic volumes, significantly extending the dynamic range without incurring prohibitively large computational expense.
To demonstrate we presented results for the galaxy stellar mass function (GSMF), the star formation rate distribution function (SFRF) and the star-forming sequence (SFS: SFR versus $M_\star$).
Our findings are as follows:

\begin{itemize}
	\item The \flares\ GSMF exhibits a clear double-Schechter shape up to $z = 10$.
				Fits assuming this form show an increasing normalisation, shallower low-mass slope and higher characteristic turnover mass with decreasing redshift.
				The GSMF is in good agreement with observational constraints at all redshifts up to $z = 8$, at which point there is some tension at the knee of the distribution.
    		The normalisation, and to a lesser extent the shape, of the GSMF shows a strong environmental dependence (i.e.~bias).
  \item The SFRF also exhibits a clear double-Schechter shape in the high-SFR regime.
				As for the GSMF, the normalisation increases and the low-mass slope decreases with decreasing redshift; however the characteristic turnover mass varies only weakly with redshfit.
  			There is a mild tension with observational results, which tend to more closely resemble power law-like distributions.
				The SFRF shape and normalisation shows a similar environmental dependence to the GSMF.
  \item The SFS shows no obvious dependence on environment.
				The low-mass slope is relatively invariant with redshift, whereas the high mass slope decreases with decresing redshift.
				The characteristic turnover mass increases slowly with decreasing redshift, and the normalisation decreases by about a factor of 3 between redshifts 10 and 5.
				There is reasonably good agreement with observational constraints at $z = 5-6$.
\end{itemize}

Upcoming space based observatories, such as JWST, \euclid\ and \romanst\ will provide further probes of the GSMF and SFRF up to $z = 10$.
The large volumes probed by \euclid\ and \romanst\ in particular will provide stronger constraints on those extreme galaxies that populate the high-mass\,/\,high-SFR tails of each distribution.
Our weighting scheme provides a means of testing the latest, high resolution hydrodynamic simulations against such constraints.  We will also be able to test the impact of cosmic variance on these large surveys.

\section*{Acknowledgements}
We wish to thank the anonymous referee for detailed comments and suggestions that improved this paper.
We also wish to thank Scott Kay and Adrian Jenkins in particular for their invaluable help getting up and running with the \eagle\ resimulation code.
Thanks also to Rob Crain, Rachana Bhatawdekar, Daniel Ceverino and Kristian Finlator for helpful suggestions and discussions.
Finally, we  thank  the \eagle\ team  for  their  efforts  in  developing the \eagle\ simulation  code.
We also wish to acknowledge the following open source software packages used in the analysis: \textsf{scipy} \citep{2020SciPy-NMeth}, \textsf{Astropy} \citep{robitaille_astropy:_2013}, \textsf{matplotlib} \citep{Hunter:2007} and \textsf{Py-SPHViewer} \citep{alejandro_benitez_llambay_2015_21703}.

This work used the DiRAC@Durham facility managed by the Institute for Computational Cosmology on behalf of the STFC DiRAC HPC Facility (www.dirac.ac.uk).
The equipment was funded by BEIS capital funding via STFC capital grants ST/K00042X/1, ST/P002293/1, ST/R002371/1 and ST/S002502/1, Durham University and STFC operations grant ST/R000832/1.
DiRAC is part of the National e-Infrastructure.
Much of the data analysis was undertaken on the {\sc Apollo} cluster at the University of Sussex.

PAT 
acknowledges support from the Science and Technology Facilities Council (grant
number ST/P000525/1).
CCL acknowledges support from the Royal Society under grant RGF/EA/181016.
APV acknowledges the support of of his PhD studentship from UK STFC DISCnet.

\section*{Data Availability}
The data underlying this article (stellar masses and star formation rates between $z = 5-10$) are available at \href{https://flaresimulations.github.io/data.html}{flaresimulations.github.io/data}.
All of the codes used for the data analysis are public and available at \href{https://github.com/flaresimulations}{github.com/flaresimulations}.


\bibliographystyle{mnras}
\bibliography{resims,custom}


\appendix

\section{Selected regions}
\label{sec:regions}

\tab{regions} lists the regions selected from the parent volume for resimulation.

\begin{table}
	\centering
	\caption{Regions selected from the parent volume for resimulation.
	We provide their positions within the parent volume, their overdensity $\delta$ as defined by \eq{1}, their \textit{rms} overdensity $\sigma$, and weights, $f_j$, calculated as per \sec{method:weighting}
	}
	\label{tab:regions}
	\begin{tabular}{ccccc} 
		\hline
		index & (x, y, z)/(\cMpch) & $\delta$ & $\sigma$ & $f_j$\\
		\hline
		0 &   (623.5, 1142.2, 1525.3)  &  0.970 &  5.62 &  0.000027 \\
		1 &   (524.1, 1203.6, 1138.5)  &  0.918 &  5.41 &  0.000196 \\
		2 &   (54.2, 1709.6, 571.1)    &  0.852 &  5.12 &  0.000429 \\
		3 &   (153.6, 1762.0, 531.3)   &  0.849 &  5.11 &  0.000953 \\
		4 &   (39.8, 1686.1, 1850.6)   &  0.846 &  5.09 &  0.000444 \\
		5 &   (847.6, 1444.0, 1062.6)  &  0.842 &  5.07 &  0.000828 \\
		6 &   (1198.2, 135.5, 1375.3)  &  0.841 &  5.07 &  0.000666 \\
		7 &   (1012.0, 1514.4, 1454.8) &  0.839 &  5.06 &  0.001178 \\
		8 &   (591.0, 359.6, 1610.2)   &  0.839 &  5.06 &  0.000265 \\
		9 &   (746.4, 820.5, 945.2)    &  0.833 &  5.03 &  0.001029 \\
		10 &  (1181.9, 1171.1, 974.1)  &  0.830 &  5.02 &  0.000387 \\
		11 &  (38.0, 670.5, 47.0)      &  0.829 &  5.02 &  0.000719 \\
		12 &  (1989.7, 368.7, 2076.5)  &  0.828 &  5.01 &  0.000668 \\
		13 &  (1659.0, 1306.6, 760.8)  &  0.824 &  4.99 &  0.000488 \\
		14 &  (57.8, 883.7, 2098.2)    &  0.821 &  4.98 &  0.001190 \\
		15 &  (609.0, 2018.6, 115.7)   &  0.820 &  4.98 &  0.000757 \\
		16 &  (122.9, 1124.1, 1304.8)  &  0.616 &  4.00 &  0.003738 \\
		17 &  (1395.2, 415.7, 1575.9)  &  0.616 &  4.00 &  0.004678 \\
		18 &  (128.3, 216.9, 258.4)    &  0.431 &  3.00 &  0.009359 \\
		19 &  (1400.6, 1686.1, 806.0)  &  0.431 &  3.00 &  0.012324 \\
		20 &  (699.4, 1760.2, 1725.9)  &  0.266 &  2.00 &  0.029311 \\
		21 &  (1951.8, 2022.3, 1709.6) &  0.266 &  2.00 &  0.027954 \\
		22 &  (755.4, 1122.3, 867.5)   &  0.121 &  1.00 &  0.057876 \\
		23 &  (516.9, 325.3, 603.6)    &  0.121 &  1.00 &  0.062009 \\
		24 &  (937.9, 1382.5, 1077.1)  & -0.007 &  0.00 &  0.074502 \\
		25 &  (1675.3, 1492.8, 1335.5) & -0.007 &  0.00 &  0.080377 \\
		26 &  (1270.5, 518.7, 862.0)   & -0.121 & -1.00 &  0.063528 \\
		27 &  (242.2, 1881.3, 1624.7)  & -0.121 & -1.00 &  0.058231 \\
		28 &  (1454.8, 1720.5, 1608.4) & -0.222 & -2.00 &  0.034467 \\
		29 &  (430.1, 296.4, 359.6)    & -0.222 & -2.00 &  0.024216 \\
		30 &  (1733.1, 1097.0, 1060.8) & -0.311 & -3.00 &  0.012087 \\
		31 &  (1821.7, 947.0, 1431.3)  & -0.311 & -3.00 &  0.013127 \\
		32 &  (1913.8, 1033.7, 45.2)   & -0.066 & -0.50 &  0.064280 \\
		33 &  (2009.6, 2024.1, 1693.4) & -0.066 & -0.50 &  0.066277 \\
		34 &  (339.8, 934.3, 1646.4)   & -0.007 &  0.00 &  0.076001 \\
		35 &  (1693.4, 914.5, 1977.1)  & -0.007 & -0.00 &  0.076486 \\
		36 &  (778.9, 900.0, 1866.8)   &  0.055 &  0.50 &  0.070408 \\
		37 &  (1790.9, 1239.7, 1765.6) &  0.055 &  0.50 &  0.062451 \\
		38 &  (2078.3, 77.7, 141.0)    & -0.479 & -5.29 &  0.002721 \\
		39 &  (818.7, 110.2, 1628.3)   & -0.434 & -4.61 &  0.003366 \\
		\hline
	\end{tabular}
\end{table}

\section{The impact of cutting passive galaxies from the star forming sequence}
\label{sec:ssfr_cut}

\begin{figure}
	\includegraphics[width=\columnwidth]{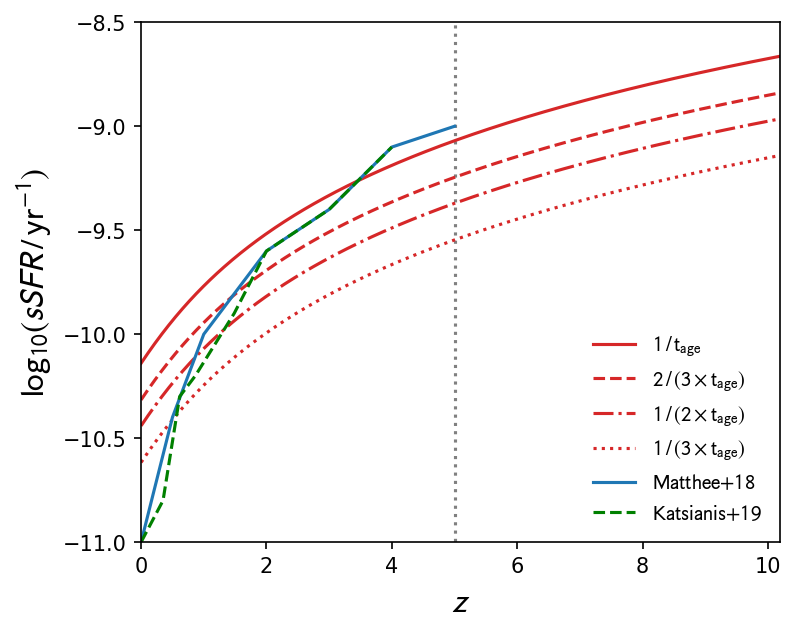}
    \caption{Evolving sSFR cuts for passive galaxies. Cuts used in \protect\cite{katsianis_evolving_2019,matthee_origin_2019} shown for comparison.
    }
    \label{fig:ssfr_cut}
\end{figure}

In \sec{results:sfs} we showed the SFS assuming no cut for passive galaxies.
We now briefly explore the impact of applying an evolving cut in specific star formation rate (sSFR), and how this impacts the SFS.
We employ an sSFR cut that excludes those galaxies whose current star formation is insufficient to double the mass of the galaxy within twice the current age of the Universe,
\begin{align}
\mathrm{sSFR} > \frac{1}{2 \times t_{\mathrm{age}}} \;\;,
\end{align}
which leads to an evolving threshold for quiescence with redshift, shown in \fig{ssfr_cut}.
Using this cut, we exclude 979 galaxies at $z = 5$ (out of a total of 32824 with stellar masses above $10^{8} \, M_{\odot}$).

\begin{figure}
	\includegraphics[width=\columnwidth]{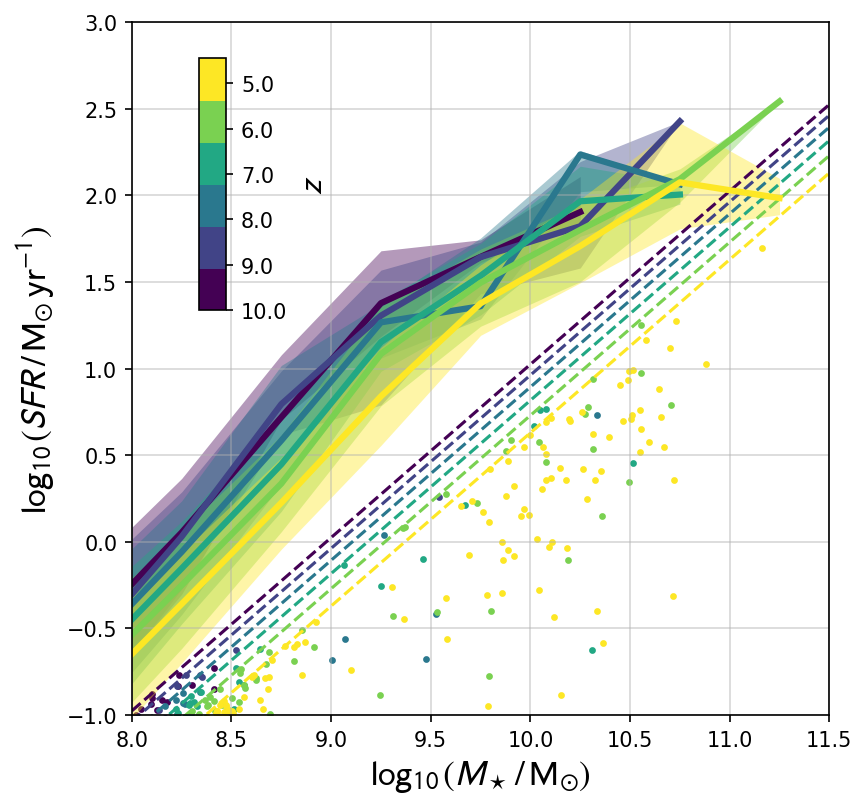}
    \caption{SFS assuming a sSFR cut for passive galaxies.
		These cuts are shown by the coloured dashed lines at each redshift.
		Points show individual passive galaxies that satisfy the cut.
		The relations are essentially identical to those without a passive galaxy cut.
    }
    \label{fig:sfs_all_cut}
\end{figure}

\fig{sfs_all_cut} shows the SFS assuming this cut.
There is almost no difference between this relation and that shown in \fig{sfs_all}.
We tested using different thresholds (mass multiples of $\times \frac{3}{2}$ and $\times 3$) and found that all our results are insensitive to the multiple of mass chosen.
Observations typically use UVJ colour to discriminate quiescent objects \citep[e.g.][]{whitaker_newfirm_2011}; at $z \sim 2$, this leads to a similar threshold for quiescence as a sSFR cut \citep{fang_demographics_2018}.

\section{Fitted distribution functions}
\label{sec:fitting}

\begin{figure*}
	\includegraphics[width=0.7\textwidth]{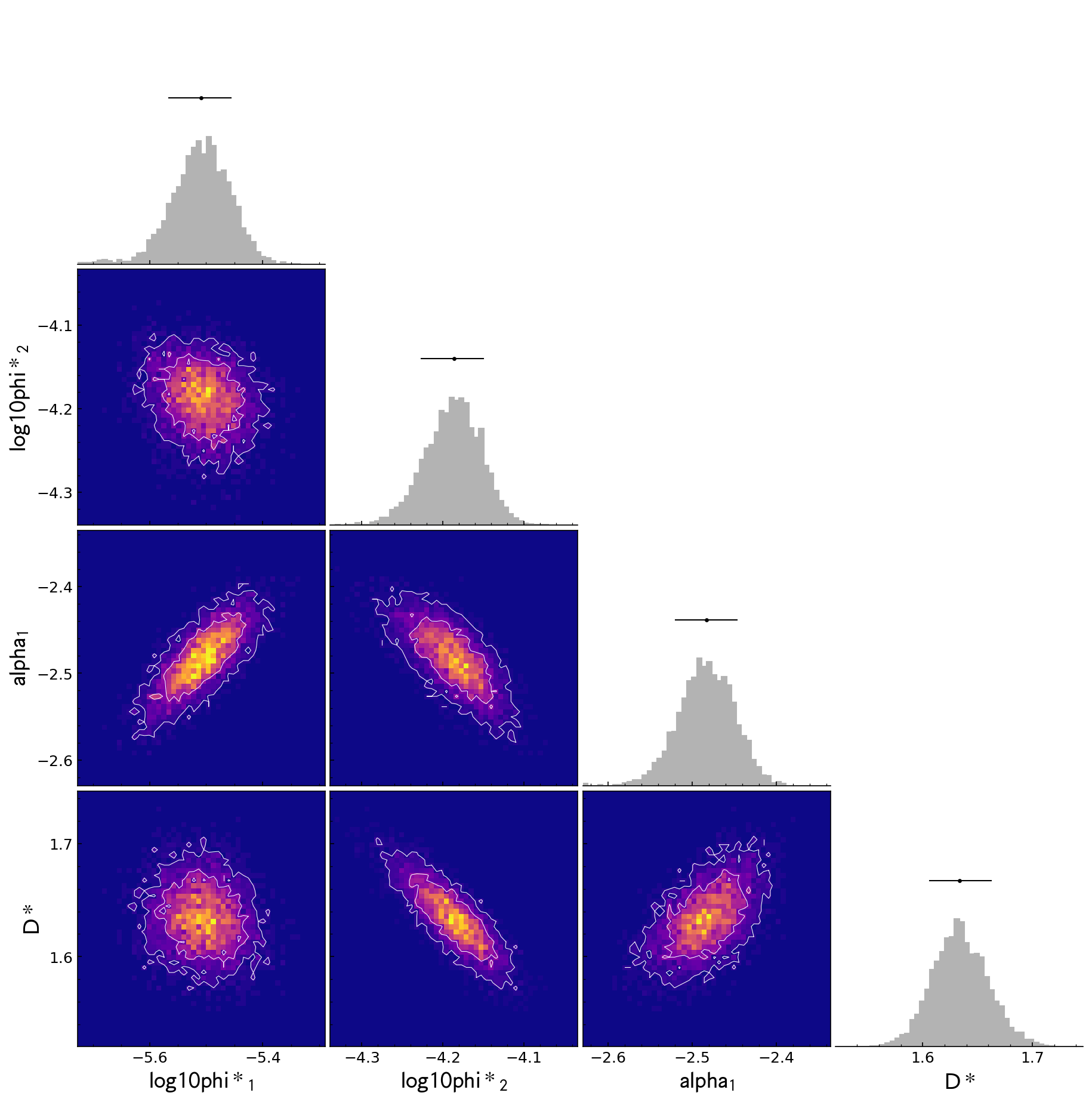}
    \caption{Posteriors from the Galaxy Stellar Mass Function fit at $z = 7$. $\alpha_2$ is fixed at -1 and is not shown.
    }
    \label{fig:posteriors}
\end{figure*}

\tab{schechter_params} and \ref{tab:sfrf_schechter_params} show double-Schechter fit parameters to the GSMF and SFRF.
We use \textsf{FitDF}, a python module for fitting arbitrary distribution functions using Markov Chain Monte Carlo (MCMC).
\textsf{FitDF} is built around the popular \textsf{emcee} package \citep[v3.0,][]{foreman-mackey_emcee_2013}.
The code can be found at \url{https://github.com/flaresimulations/fitDF}.

A Poisson form of the likelihood is typically used for distribution function analyses in Astronomy due to the relatively small number of observations.
Due to our resimulation approach we cannot use this form of the likelihood, since the number counts obtained from the composite approach, scaled to the size of the parent box volume, significantly underestimate the errors.
Instead, we use a Gaussian form for the likelihood,
\begin{align}
\mathrm{log}(\mathcal{L}) = - \frac{1}{2} \, \left[ \sum_{i} \, \frac{ (N_{i,\mathrm{obs}} - N_{i,\mathrm{exp}})^2 }{\sigma_{i}^{2}} \, + \mathrm{log}(\sigma_{i}^{2}) \right] \;\;,
\end{align}
where the subscript $i$ represents the bin of the property being measured, $N_{i,\mathrm{obs}}$ is the inferred number of galaxies using the composite number density multiplied by the parent box volume, $N_{i,\mathrm{exp}}$ is the expected number from the model, and $\sigma_i$ is the error estimate.
Using this form, $\sigma$ can be explicitly provided from the resimulated number counts, $\sigma_{i} = N_{i,\mathrm{obs}} / \sqrt{n_{i,\mathrm{obs}}}$, where $n_{i,\mathrm{obs}}$ is the number counts in bin $i$ from the resimulations.

We use flat uniform priors in $\mathrm{log_{10}}(D^*)$, $\alpha_1$, $\mathrm{log_{10}}(\phi^*_1)$ and $\mathrm{log_{10}}(\phi^*_2)$.
We fix $\alpha_2 = -1$ by setting a narrow top-hat prior around this value.
We run chains of length $10^{4}$, then calculate the autocorrelation time, $\tau$, on these chains \citep{goodman_ensemble_2010}.
We use $\tau$ to estimate the burn-in ($\tau \times 4$) and thinning ($\tau / 2$) on our chains.\footnote{The chains for each fit are available at \url{https://flaresimulations.github.io/flares/data.html}.}
Example posteriors for each parameter in a fit to the $z = 7$ GSMF are shown as a corner plot in \fig{posteriors}.

\tab{sfs_params} shows the piecewise-fits to the SFS; the fitting procedure is described in \sec{results:sfs}.

\begin{table}
	\centering
	\begin{tabular}[t]{cccccc}
		\hline
		z & $x_{0} + 9.7$ & $\alpha_{1}$ & $\alpha_{2}$ & $\beta$ \\
		\hline
		5  & 9.60 & 1.23 & 0.62 & 1.42 \\
		6  & 9.45 & 1.27 & 0.70 & 1.60 \\
		7  & 9.35 & 1.31 & 0.72 & 1.76 \\
		8  & 9.20 & 1.33 & 0.80 & 1.90 \\
		9  & 9.16 & 1.31 & 0.91 & 1.93 \\
		10 & - & 1.24 & - & - \\
		\hline
	\end{tabular}
	\caption{Best fitting two-part piecewise-linear fits to the star-forming sequence.}
	\label{tab:sfs_params}
\end{table}

\begin{table*}
	\centering
	\begin{tabular}[t]{cccccc}
		\hline
		z &  M$^{*}$ & log$_{10}$($\phi^{*}_{1}$\,/(Mpc$^{-3}$ dex$^{-1}$)) & log$_{10}$($\phi^{*}_{2}$\,/(Mpc$^{-3}$ dex$^{-1}$)) & $\alpha_1$ \\
		\hline
		10 &            $9.117_{ - 0.045 }^{ +0.041 }$ &            $-6.557_{-0.197}^{+0.188 }$ &            $-4.871_{-0.07}^{+0.065 }$ &            $-3.542_{-0.206}^{+0.193}$ \\
		9 &            $9.488_{ - 0.044 }^{ +0.036 }$ &            $-6.372_{-0.112}^{+0.116 }$ &            $-4.832_{-0.057}^{+0.056 }$ &            $-3.07_{-0.077}^{+0.076}$ \\
		8 &            $9.577_{ - 0.041 }^{ +0.039 }$ &            $-5.904_{-0.08}^{+0.081 }$ &            $-4.565_{-0.058}^{+0.059 }$ &            $-2.83_{-0.048}^{+0.065}$ \\
		7 &            $9.831_{ - 0.035 }^{ +0.039 }$ &            $-5.443_{-0.054}^{+0.051 }$ &            $-4.374_{-0.059}^{+0.052 }$ &            $-2.515_{-0.032}^{+0.03}$ \\
		6 &            $10.089_{ - 0.035 }^{ +0.029 }$ &            $-5.057_{-0.047}^{+0.036 }$ &            $-4.156_{-0.046}^{+0.05 }$ &            $-2.293_{-0.023}^{+0.019}$ \\
		5 &            $10.326_{ - 0.02 }^{ +0.019 }$ &            $-4.686_{-0.024}^{+0.023 }$ &            $-3.942_{-0.034}^{+0.033 }$ &            $-2.11_{-0.011}^{+0.012}$ \\
		\hline
	\end{tabular}
	\caption{Best fitting double-Schechter function parameter values for the Galaxy Stellar Mass Function. $\alpha_{2}$ is fixed at $-1$.} \label{tab:schechter_params}
\end{table*}

\begin{table*}
	\centering
	\begin{tabular}[t]{cccccc}
		\hline
		z &  SFR$^{*}$ & log$_{10}$($\phi^{*}_{1}$\,/(Mpc$^{-3}$ dex$^{-1}$)) & log$_{10}$($\phi^{*}_{2}$\,/(Mpc$^{-3}$ dex$^{-1}$)) & $\alpha_1$ \\
		\hline
		5 &            $1.402_{ - 0.067 }^{ +0.049 }$ &            $-6.525_{-0.123}^{+0.142 }$ &            $-5.022_{-0.069}^{+0.07 }$ &            $-2.978_{-0.074}^{+0.071}$ \\
		5 &            $1.359_{ - 0.044 }^{ +0.036 }$ &            $-5.941_{-0.093}^{+0.093 }$ &            $-4.645_{-0.058}^{+0.058 }$ &            $-2.772_{-0.06}^{+0.064}$ \\
		5 &            $1.433_{ - 0.028 }^{ +0.032 }$ &            $-5.639_{-0.066}^{+0.059 }$ &            $-4.431_{-0.058}^{+0.049 }$ &            $-2.62_{-0.045}^{+0.051}$ \\
		5 &            $1.633_{ - 0.027 }^{ +0.03 }$ &            $-5.509_{-0.057}^{+0.052 }$ &            $-4.186_{-0.04}^{+0.036 }$ &            $-2.482_{-0.038}^{+0.036}$ \\
		5 &            $1.684_{ - 0.015 }^{ +0.015 }$ &            $-5.059_{-0.039}^{+0.041 }$ &            $-3.907_{-0.026}^{+0.024 }$ &            $-2.307_{-0.025}^{+0.026}$ \\
		5 &            $1.755_{ - 0.012 }^{ +0.011 }$ &            $-4.68_{-0.033}^{+0.033 }$ &            $-3.644_{-0.02}^{+0.02 }$ &            $-2.139_{-0.019}^{+0.02}$ \\
		\hline
	\end{tabular}
	\caption{Best fitting double-Schechter function parameter values for the Star Formation Rate distribution function. $\alpha_{2}$ is fixed at $-1$.}
	\label{tab:sfrf_schechter_params}
\end{table*}




\bsp	
\label{lastpage}
\end{document}